\documentclass[11pt]{article}

\usepackage{graphicx}
\usepackage[toc,page]{appendix}
\usepackage{multicol} 

\usepackage{color}
\usepackage{latexsym}
\usepackage{bm}

\definecolor{rosso}{cmyk}{0,1,1,0.4}
\definecolor{rossos}{cmyk}{0,1,1,0.55}
\definecolor{rossoc}{cmyk}{0,0.5,1,0.2}
\definecolor{blu}{cmyk}{1,1,0,0.3}
\definecolor{blus}{cmyk}{1,1,0,0.6}
\definecolor{blucc}{cmyk}{1,0.4,0.2,0}
\definecolor{viola}{cmyk}{0,1,0,0.6}
\definecolor{viola2}{cmyk}{0,1,0.2,0.6}
\definecolor{verde}{cmyk}{0.92,0,0.59,0.25}
\definecolor{verdec}{cmyk}{0.92,0,0.59,0.15}
\definecolor{verdes}{cmyk}{0.92,0,0.59,0.4}
\font\tenrsfs=rsfs10 at 12pt
\font\sevenrsfs=rsfs7
\font\fiversfs=rsfs5
\newfam\rsfsfam
 
\textfont\rsfsfam=\tenrsfs
\scriptfont\rsfsfam=\sevenrsfs
\scriptscriptfont\rsfsfam=\fiversfs
\def\mathscr#1{{\fam\rsfsfam\relax#1}}

\def\circa#1{\,\raise.3ex\hbox{$#1$\kern-.75em\lower1ex\hbox{$\sim$}}\,}

\usepackage{amsmath,latexsym,amssymb,color,hyperref,graphicx}
\newcommand{\eq}[1]{(\ref{#1})}
\newcommand{\be}{\begin{equation}}
\newcommand{\ee}{\end{equation}}
\newcommand{\bea}{\begin{eqnarray}}
\newcommand{\ena}{\end{eqnarray}}
\newcommand{\no}{\noindent}
\newcommand{\nb}{\nonumber}

\renewcommand\l{\lambda}
\renewcommand\o{\omega}  
\renewcommand\a{\alpha}
\renewcommand\b{\beta}

\renewcommand\l{\ensuremath{\lambda}}

\renewcommand\L{\ensuremath{\Lambda}}
\newcommand\m{\ensuremath{\mu}}
\renewcommand\k{\ensuremath{\kappa}}

\newcommand\n{\ensuremath{\nu}}

\newcommand{\de}{\partial}

\renewcommand\l{\ensuremath{\lambda}}

\newcommand{\ba}{\begin{eqnarray}}
\newcommand{\ea}{\end{eqnarray}}
\newcommand{\plm}{M_{\text{Pl}}} 
 
\makeatletter
\def\ps@mine{%
    \def\@oddfoot{\hfil\thepage\hfil}\let\@evenfoot\@oddfoot
    \let\@oddhead\@evenhead%
    \let\@mkboth\@gobbletwo
    \let\sectionmark\@gobble
    \let\subsectionmark\@gobble
    }
\renewcommand\section{\@startsection {section}{1}{\z@}%
                                   {-3.5ex \@plus -1ex \@minus -.2ex}%
                                   {2ex \@plus.2ex}%
                                   {\normalfont\large\sffamily\bfseries}}
\renewcommand\subsection{\@startsection {subsection}{1}{\z@}%
                                   {-3.5ex \@plus -1ex \@minus -.2ex}%
                                   {2ex \@plus.2ex}%
                                   {\normalfont\sffamily\bfseries}}
\makeatother

\numberwithin{equation}{section}

\begin{document}
\thispagestyle{empty}
\vspace*{-2.5cm}
\begin{minipage}{.45\linewidth}
\begin{flushleft}                           
\end{flushleft} 
\end{minipage}
\vspace{2.5cm}

\begin{center}
{\huge\sffamily\bfseries 
Intrinsic Entropy Perturbations from  the Dark Sector
 }
 \end{center}
 
 \vspace{0.5cm}
 
 \begin{center} 
 {\sffamily\bfseries \large  Marco Celoria}$^{a}$,  
 {\sffamily\bfseries \large Denis Comelli}$^b$,
  {\sffamily\bfseries \large Luigi Pilo$^{c,d}$}\\[2ex]
  {\it
$^a$ Gran Sasso Science Institute (INFN)\\Via Francesco Crispi 7,
L'Aquila, I-67100\\\vspace{0.1cm}
$^b$INFN, Sezione di Ferrara,  I-35131 Ferrara, Italy\\\vspace{0.1cm}
$^c$Dipartimento di Fisica, Universit\`a dell'Aquila,  I-67010 L'Aquila, Italy\\\vspace{0.1cm}
$^d$INFN, Laboratori Nazionali del Gran Sasso, I-67010 Assergi, Italy\\\vspace{0.3cm}
{\tt marco.celoria@gssi.infn.it}, 
 {\tt comelli@fe.infn.it}, {\tt luigi.pilo@aquila.infn.it}
}
\end{center}
   
\vspace{0.7cm}

\begin{center}
{\small \today}
\end{center}

\vspace{0.7cm}

\begin{center}
{\sc Abstract}

\end{center}
\no
Perfect fluids are modeled by using an effective field
theory approach which naturally gives a self-consistent and unambiguous
description of the intrinsic non-adiabatic contribution to pressure variations.
We study the  impact  of  intrinsic entropy perturbation on the superhorizon dynamics of
the curvature perturbation ${\cal R}$  in the dark sector. The dark sector, made of dark
matter and dark energy is described as a single perfect fluid. 
The non-perturbative vorticity's dynamics and the 
Weinberg theorem violation for perfect fluids are also studied.

 \vspace{0.5cm}

\section{Introduction}
The Universe is undergoing a phase of
accelerated expansion~\cite{Riess:1998cb,Perlmutter:1998np}. What is
is driving such an accelerated phase is presently not known and a  number of forthcoming dark energy surveys will try to shed some
light on  its nature, see for instance~\cite{Drlica-Wagner:2017tkk,Amendola:2016saw,Spergel:2015sza,Abell:2009aa}.
Besides the cosmological constant, a vast  number of
models have been proposed, ranging from proposals of   modifications of
gravity to the addition of exotic matter. Unfortunately, only a small  number of observables are
available from the surveys and discriminating among the various models
is going to be very challenging. For this reason it important to build
the simplest possible  effective description for a dark  fluid that is
able to capture all the relevant physical properties distinguishing
dark energy from  cosmological constant. Recently~\cite{ussgf,classus}
is has been proposed 
an effective description of dark sector modelled as a generic
self-gravitating medium. Such a medium is
described by the theory of four derivatively coupled scalar which can be
interpreted as comoving coordinates of the medium whose fluctuations
represent the Goldstone modes for the broken spacetime
  translations. The very same scalar fields can be viewed as
  St\"uckelberg fields that allow to restore broken 
diffeomorphisms
~\cite{Leutwyler:1993gf,Leutwyler:1996er,ArkaniHamed:2002sp,ussgf,Rubakov:2008nh,Dubovsky:2004sg}.
Such an effective field theory description has been already considered
in~\cite{Dubovsky:2005xd,Dubovsky:2011sj,Nicolis:2011cs,Ballesteros:2012kv}
for particular type of media.
Internal symmetries of the medium action determine both the
dynamical and the thermodynamical properties of the system.
 In the present study we will focus on the phenomenological
consequences of  taking dark energy as
the simplest class of self-gravitating media: perfect fluids. Though
perfect fluids are rather simple, taking  into account  a
  non-barotropic equation of state leads 
to an additional source in the growth of structure and to the dynamics of the  comoving curvature
perturbation ${\cal R}$. The effective field theory description allows
us to determine in a consistent and compact way the form of the
non-adiabatic contributions induced by pressure variations. The 
formalism is also employed to study when the Weinberg
Theorem~\cite{Weinberg:2003sw, 
Weinberg:2008zzc} is  violated.\\
The outline of the paper is the following.
 In section \ref{sect:action} we introduce  the  effective
 action which gives the non perturbative dynamical  description of  a
 generic    non-barotropic  perfect fluid, together with
 the relation among thermodynamical variables and field theory
 operators;  in relation with previous literature,
 \cite{Ellis:1989ju,Ellis:1990gi}, at non-perturbative level, the influence
of entropy on the time evolution the fluid's vorticity.  
Section \ref{sect:FRW} is devoted to linear
cosmological perturbations in a flat FRW Universe in the presence of a single entropic perfect fluid
by using its effective field theory description. 
We analyse also the conditions for the violation of Weinberg
 theorem~\cite{Weinberg:2003sw, 
Weinberg:2008zzc} in the presence of perfect fluids, confirming and
extending the known results.
Such results are extended to
multifluids in section \ref{section:multi}. The phenomenological
consequences of the entropic modes are considered for different
descriptions of the dark sector.    The
$\Lambda$CDM model described in section \ref{sect:LCDM} is compared
with different  fluid models of the dark sector in sections
 \ref{case1}, \ref{case2}, \ref{case3}.   Our conclusions are given in section \ref{sect:conc}.

\section{Perfect Fluids Action and Thermodynamics}
\label{sect:action}
Perfect fluids and  non dissipative general media can be described by using an effective
field theory approach in terms of 
four scalar fields $\Phi^A$  ($A=0,1,2,3$). The medium physical
properties are encoded in a set of  symmetries of the scalar field
action  selecting, order by order in a derivative expansion, a finite
number of operators.
Following~\cite{ussgf,usthermo} we require the Lagrangian to be invariant under:
\be
\begin{split}
\text{global shift symmetries:}&  \quad \Phi^A\to \Phi^A+f^A \, ; \\
\text{field dependent symmetries:} &\quad \Phi^0\to \Phi^0+f(\Phi^a);\quad
\Phi^a\to f^a(\Phi^b),\quad \det\left|\frac{\partial
    f^a}{\partial\Phi^b}\right|=1 \, ;\\
\text{internal rotational invariance:}& \quad \Phi^a \to {\cal M}^a_b\;\Phi^b \, , \quad {\cal M} \in \text{SO}(3) \, .  \end{split} 
\ee
The global shift symmetry requires the scalars appear in the action
only through their derivatives,  while field dependent symmetries plus
internal rotational invariance select  the following operators
\ba
b=\left[\det \left(g^{\mu \nu} \partial_\mu  \Phi^a\,\partial_\nu
    \Phi^b \right) \right]^{1/2},\qquad
u^\mu=\frac{1}{b\;\sqrt{g} }\epsilon^{\mu\alpha \beta \gamma}\;\partial_\alpha\Phi^1\;\partial_\beta\Phi^2\;\partial_\gamma\Phi^3,\qquad
Y=u^\mu\;\partial_\mu\Phi^0 \,  ;
\label{shift}
\ea
with $u^2\equiv u^\mu u_\mu = -1$.
As a result, our starting point is the most general action  given by
\be
S=\int d^4x\;\sqrt{g}\;U(b,\,Y) \, ;
\ee
that gives the  following conserved energy momentum tensor (EMT) 
\be
T_{\mu\nu}=\rho \; u_\mu\;u_\nu+(\rho + p) \; g_{\mu\nu} \, ,
\label{emt}
\ee
with   a perfect fluid form where~\footnote{We have introduced the notation $\frac{\de U}{\de Y}
  =U_Y$, $\frac{\de U}{\de b}
  =U_b$ and $\frac{\de^2 U}{\de Y\de Y}
  =U_{Y^2}$, etc.}  
\ba\label{rp}
\rho=-U+Y\;U_Y,\qquad
p=U-b\;U_b \, .
\ea
The shift symmetries (\ref{shift}) lead to the existence of two
conserved currents~\footnote{Actually there more conserved currents
  that will not be needed here; see~\cite{ussgf} for a complete analysis.}
\ba\label{trt}
&& \nabla^\mu J_\mu = \nabla^\mu  (b\,u_\mu)=\dot b+\theta\;b=0 \, ;\\ \label{trt1}&&
 \nabla^\mu {\cal J}_\mu = \nabla^\mu  (U_Y\,u_\mu)=\dot U_Y+\theta\;U_Y=0
 \, ;
\ea
where $\dot f=u^\nu\nabla_\nu f$ is the Lie derivative of $f$ along $u^\m$.
It also follows from (\ref{trt}, \ref{trt1}) that  the ratio
$\sigma\equiv \frac{U_Y}{b}$ is conserved, indeed
\be\label{sc}
u^\mu\nabla_\mu\,\sigma\equiv\dot \sigma=0 \, .
\ee
The equations (\ref{trt}, \ref{trt1}) are equivalent to the projection of the
EMT conservation \eq{emt} along $u^\mu$. 

One can also relate the operators $b$ and
$Y$ with thermodynamical variables of the fluid, namely the entropy
density  $s$,   the chemical potential  $\mu$ and the temperature $T$.
The  basic  input is the first
principle $d\rho=T\;ds+\mu\;dn$
and the Euler Relation 
$\rho=T\;s-p+\mu\;n$ or equivalently the Gibbs-Duhem equation
$dp=s\,dT-n\,d\mu$. 
As shown in ~\cite{usthermo}, two consistent thermodynamical
interpretation relating the operators $b$ and $Y$ and
thermodynamical variables exist, namely  
\ba\label{ch1}
&&T=Y,\qquad s=U_Y,\qquad n=b\qquad \mu=-U_b \, ;
\\
&& T=-U_b,\qquad s=b,\qquad n=U_Y\qquad \mu=Y \, .
\ea
where we neglect  a constant normalisation factor for each equality.
For the rest of the paper we will consider the relationship \eq{ch1}. 
From \eq{ch1},  the conserved quantity $\sigma=\frac{s}{n}$ (see
eq.\eq{sc}) is the entropy per particle and it is always conserved for
perfect fluids.
  The current
\be
J_\mu= b \;u_\m 
\ee
is always conserved independently from the equations of motion of the
scalar fields, with $b=n$ representing the number particle density.
\\
The hydrodynamic equations for a perfect fluid  are equivalent to
the EMT and current conservation
\be
\nabla^\nu T_{\mu\nu}=0 \, ,\qquad \nabla^\mu J_\mu=0 \, .
\ee
Indeed, projecting the EMT conservation equation along $u^\mu$ and on
an  orthogonal direction 
by using the projector  $h_\mu^\nu=\delta_\mu^\nu+u_\mu\,u^\nu$,
we have that 
\ba\label{eqpf}
\dot \rho+\theta\;(p+\rho)=0,\qquad D_\mu\,p=-(p+\rho) a_\mu\;
\ea
where
\be
\theta=\nabla_\mu\,u^\mu \, ,
 \qquad  D_\mu=h_\mu^\nu\,\nabla_\nu \, , \qquad a_\mu = u^\nu
 \nabla_\nu u_\mu \, .
\ee
The  study of the { differential} properties of the pressure
 allows to single out the non adiabatic contribution $\mathbf{\Gamma}
$ ~\cite{Kodama:1985bj}. 
 Indeed, knowing $p$ as function of $b$ and $Y$  by eq \eq{rp} and  
using  (\ref{ch1}) we can express $p$ also as a function of $\rho$ and
$\sigma$, thus
\be\label{dp}
 dp  =\left.\frac{\partial p}{\partial \rho}\right|_{\sigma}\;d
\rho+\left.\frac{\partial p}{\partial \sigma}\right|_{\rho}\; d \sigma\;\equiv\; c_s^2\; d \rho+ \mathbf{\Gamma}
\, ;
\ee
$c_s^2$ is the adiabatic sound speed. 
Notice that the definition (\ref{dp}) of $\mathbf{\Gamma}$ is general and non perturbative.

The EFT description together with the thermodynamical
dictionary (\ref{ch1}) allow us to compute explicitly
$\mathbf{\Gamma}$.  From the knowledge of $p$, $\rho$ and $
\sigma$ as function of $b$ and $Y$, the 1-forms $d\,b$ and $d\,Y$ can be
expressed in terms of $d\,\rho$ and $d\,\sigma$. Actually, $c_s^2$ can be
computed in a simpler way by   contracting the 1-form $d\,p$ with $u^\m$ to
get 
\be
d\,p(u) = u^\mu \de_\mu p= \dot{p} = c_s^2 \;\dot \rho +
\left.\frac{\partial p}{\partial \sigma}\right|_{\rho}  \, \dot \sigma
 =  c_s^2 \;\dot \rho  \, ,  \qquad {\rm being}\;\;\;\dot \sigma=0 \, ;
\ee
thus 
\be
c_s^2 = \left.\frac{\partial p}{\partial \rho}\right|_{\sigma}=
\frac{\dot{p}}{\dot{\rho}} = \frac{p_b \, \dot{b} + p_Y \,
  \dot{Y}}{\rho_b \, \dot{b} + \rho_Y \, \dot{Y}} =
\frac{(U_Y-b\;U_{bY})^2-b^2\;U_{b^2}\;U_{Y^2}}{{  U_{Y^2}}\;(\rho+p)} \, ;
 \label{cs}
\ee
where we have used (\ref{trt}) and (\ref{trt1}) to eliminate $\dot{b}$
and $\dot{Y}$. For ${\mathbf{ \Gamma}}$ we get
\be
 \mathbf{\Gamma}= \left.\frac{\partial p}{\partial
     \sigma}\right|_{\rho}  \, d\,\sigma \equiv c_\rho^2\;d\,\sigma= b\;Y\;(c_b^2-c_s^2)\, d\,\sigma  \, . \label{dpi}
\ee
with
\be
c_\rho^2\; = \left.\frac{\partial p}{\partial
     \sigma}\right|_{\rho} =b\;Y\;(c_b^2-c_s^2)\;\qquad c_b^2=\left.\frac{\partial p}{\partial
    \rho}\right|_{b } =\left.\frac{\partial p}{\partial
    \rho}\right|_n= \frac{U_Y-b\;U_{bY}}{Y\;U_{Y^2}} \, . 
\label{cb}
\ee
From
the conservation of ${\cal J}_\mu$ one can derive an evolution
equation for $Y$; expanding (\ref{trt1}) and using (\ref{trt}) to
eliminate $\dot{b}$, one gets  by using the definition (\ref{cb}) of $c_b^2$
\be
\dot{Y}=- c_b^2\;\theta\; Y \, ;
\label{Ydot}
\ee
in addition (\ref{trt1}) and  (\ref{trt}) can be rewritten as
\be
\frac{\dot{U}_Y}{U_Y} = \frac{\dot{b}}{b} \, .
\label{Ybrel}
\ee
Finally, by contracting $d\,p$ with
 the orthogonal projector $h_\m^\n$ one gets
\ba
\label{ddp}
D_\a\, p=c_s^2\;D_\a \rho +
c_\rho^2\, D_\a\, \sigma \, ;
\ea
where $D_\mu f= h_{\mu \nu} \, \nabla^\nu f$ is the projected
covariant derivative acting on a scalar function $f$.
Once the perfect fluid is specified giving $U(b,Y)$, one can
compute $c_s^2$, $c_b^2$ and $c_\rho^2$,    then $\mathbf{\Gamma}
$ is known in a fully non
perturbative way.  
As we will see the knowledge of $\mathbf{\Gamma}
$ is crucial
to study the dynamics of the gravitational potential in a perturbed  Friedman Lemaitre Robertson Walker (FLRW)
Universe.

In~\cite{lv1,lv2} it was shown that it is possible to give a non
perturbative gauge invariant definition of 
entropic variables, defining   appropriate spatial-gradient quantities,   leading to   simple geometric nonlinear conserved quantities for a perfect fluid
\ba
\zeta_a\equiv
D_a\a-\frac{\dot\a}{\dot\rho}\;D_a\rho=D_a\a+\frac{D_a\rho}{3\;(\rho+p)}\;,\qquad
\theta=3\;\dot\a \, .
\ea
The ``evolution" equation for the $\zeta_a$ vector is given by the
following Lie derivative obtained using the equations of motion of the
perfect fluid \eq{eqpf} (the Lie derivative of a 1-form is given by
$ {\cal L}_u (\zeta_a)=\;\zeta_a'+ \zeta_b\;\nabla_a\;u^b$)
\ba\label{eqz}
{\cal L}_u(\zeta_a)=-\frac{\theta}{3\;(\rho+p)}\left( D_a p-\frac{\dot
    p}{\dot \rho}\;D_a\rho\right) \, .
\ea
This result is valid for any spacetime geometry and does not depend on Einstein's equations.
In the cosmological context, $\alpha$ can be interpreted as a non-linear generalisation, according to an observer following the fluid, of the number of e-folds of the scale factor.
On scales larger than the Hubble radius, the above definitions are equivalent to the non-linear curvature perturbation on uniform density hypersurfaces \cite{lv1}.
From \eq{ddp} and \eq{dpi} we can write
\ba\label{Lz}
 {\cal L}_u(\zeta_a)=-\frac{\theta\;c_\rho^2}{3\;(\rho+p)}\;D_a\sigma
=-\frac{\theta\;n\;T\,(c_b^2-c_s^2)}{3\;(\rho+p)}\;D_a\sigma
\quad {\rm with}\quad \dot\sigma=0 \, .
\ea
For isentropic perturbations, $D_\a\sigma=0$, \eq{eqz} vanishes identically   and 
the above guarantees that ${\cal L}_u(\zeta_a)=0$, i.e. 
$\zeta_a$ is a conserved quantity in the isentropic/barotropic case on all scales and
at all perturbative orders.
  By defining the vorticity tensor $\Omega_{\a \b}$, the enthalpy $h$
and the enthalpy vector $w_\a$ as 
\be
 h=\frac{p+\rho}{n} \, , \qquad w_\a=h\;u_\a \, , \qquad
 \Omega_{\a\b}\equiv\nabla_\b\;w_\a-\nabla_\a\;w_\b
\ee
we note that the spatially projected EMT conservation   equation  can be also rewritten in form
\be
\label{cl}
\Omega_{\a\b}\,u^\b=T\;D_\a\sigma
\ee
as proposed by  Carter  and  Lichnerowicz~\cite{lichne, Carter:1987qr},
 As a consequence, if a perfect fluid has zero vorticity
tensor, i.e $\Omega_{\mu\nu}=0$, then   equation \eq{cl} immediately implies that irrotational perfect fluids are  also  isentropic. 
 From the  definition of $\Omega_{\a\b}$ it follows that
if the vorticity is zero, then the velocity of a relativistic
(isentropic) perfect fluid can be expressed as the gradient of a
  velocity potential function.
More specifically, the enthalpy current  can be expressed, at least locally, as the gradient of a potential $F$,
$w_\m=h\;u_\mu=\nabla_\mu\;F$.
  Of course, an  isentropic fluid  does not necessarily have zero vorticity.
Equation \eq{Lz} can also be rewritten in the form
\be\label{}
 {\cal L}_u(\zeta_\a)
=-\frac{\theta\;(c_b^2-c_s^2)}{h}\; \Omega_{\a\b}\;u^\b \, .
\ee
A physical observable that is particularly sensitive to entropic
perturbations is the  kinematic vorticity tensor and the  kinematic vorticity vector,
defined as
\be
\omega_{\mu\nu}=\frac{1}{2}\;\left(D_{\mu}u_{\nu}-D_{\nu}u_{\mu}\right)
,\quad
\omega^\mu=\frac{1}{2}\;\epsilon^{\mu\a\b\gamma}\;u_\gamma\;\omega_{\a\b}
\ee
whose non-perturbative time evolution equation is given by \cite{Ellis:1989ju,Ellis:1990gi}
\ba\label{ha}
h^\nu_\m\;\dot\o_\n=-\frac{2}{3}\;\theta\;\o_\m+\sigma_{\m\n}\;\o^\n-\frac{1}{2}\;{\rm
  curl} (a_\mu )\, ;
\ea
where ${\rm curl}( a_\mu)=\epsilon_{\mu\a\b }\, D^\a\,a^\b$ and $\epsilon_{\mu\a\b}=\epsilon_{\mu\a\b\gamma}\;u^\gamma$ and
${\rm curl}( D_\mu f)=-2\;\dot f\;\o_\m$.
\\
By using the following relations
\be
{\rm curl}( D_\mu p)=2\;c_s^2\;\theta\;(\rho+p)\;\o_\m \, , \qquad \qquad
\epsilon_{\mu\a\b } \;D^\a \rho\;D^\b p
= c_\rho^2\;\epsilon_{\mu\a\b } \;D^\a \rho\;D^\b \sigma \, ;
\ee
we can expand the source term of equation \eq{ha} as
\ba
{\rm curl} (a_\mu) =
-2\;c_s^2\;\theta\;(\rho+p)\;\o_\m+\frac{c_\rho^2}{(\rho+p)^2}\;\epsilon_{\mu\a\b
} \;D^\a \rho\;D^\b \sigma \, .
\ea
Thus for an entropic perfect fluid we can write
\ba\label{haf}
h^\nu_\m\;\dot\o_\n+\;\theta\;(\frac{2}{3}-
c_s^2)\;\o_\m-\sigma_{\m\n}\;\o^\n=\frac{1}{2}\;
\frac{c_\rho^2}{(\rho+p)^2}\;\epsilon_{\mu\a\b } \;D^\a \sigma\;D^\b
\rho \, ,
\ea
  where $\sigma_{\mu \nu}$ is the shear tensor. 
This equation shows that there is no source for vorticity for an adiabatic perfect fluid at any perturbative order. 
Note that around a FRW background the second term on the left is order one, the third (proportional to the shear that is at least of order two) is order three  and the source term is of order two \cite{Christopherson:2009bt,Brown:2011dn,Christopherson:2011ep,Christopherson:2014eoa}. 
Equations \eq{Lz},\eq{haf} are intrinsically non perturbative and
show the importance of the the entropy per particle in the evolution
of a perfect fluid. 
In the next section we will   move on  to study how
linear  scalar perturbations are affected by the presence of non-adiabatic perturbations
by exploiting the power  of the effective field theory formalism.

\section{Cosmology with   Perfect Non Barotropic Fluids}
\label{sect:FRW}

\subsection{FLRW Cosmology}
Let us consider a  spatially flat FLRW Universe 
\be
ds^2 = a^2 \, \eta_{\mu \nu} \;dx^\mu dx^\nu \, ,
\label{FRW}
\ee
with matter described by perfect fluid  EMT tensor (\ref{emt})
with $u^\mu=(a^{-1},0,0,0)$.
The dynamics of the scale
 factor $a$ is determined by
\be\label{emc}
\rho'+3\;{\cal H}\;(\rho+p)=0,\qquad {\cal H}^2 \equiv \left(
  \frac{a'}{a}\right)^2=\frac{a^2\, \rho}{6 \, \plm^2} \, ;
\ee 
where $'$ denotes the derivative with respect to conformal time.
 On FLRW $\dot{f} \equiv u^\mu \de_\mu f = a^{-1} \, f'$ and $\theta = 3\;
 {\cal H}$.
 It is convenient for what follows to define the equation of state $w$ of the fluid 
\be\label{pG}
  w\equiv\frac{p}{\rho}\, , \qquad 
  w'=3\;{\cal H} \;(1+w)\;(w-c_s^2) \, ;
 \ee
where  $c_s^2 \equiv \dot{p}/\dot{\rho}=p'/\rho'$ is the sound speed given by \eq{cs},
evaluated at the leading order in cosmological perturbation theory
(background level).
From \eq{trt} we can determine $b$ in terms of $a$ as
\be\label{eqb}
b
=\frac{b_0}{a^3} \, ,
\ee
which also leads to the usual scaling of number density  $n$
according to \eq{ch1} as $n=n_0/a^3$, with $n_0$ the present density
and we have chosen $a=1$ as the today's value of the scale factor.
By integrating \eq{Ybrel}, one finds that  $Y$  is algebraically   related
to $b$  by the equation
\be U_Y(b,\,Y)={\sc c}\;b \, ,
\ee
where ${\sc c}$ is a constant. Notice also that at the background
level, setting $\Phi^0\equiv \phi$ we have
$Y= \dot\Phi^0= \phi'/a^{}$. 
When $c_b^2$ is time independent, the integration of \eq{Ydot} can be
used to deduce how the fluid temperature, identified with $Y$,  scales
with the scale factor 
\be
T=\frac{T_0}{a^{3\;c_b^2}},\,\quad {\rm for} \quad c_b^2=\text{const.}
\, , 
\label{Tscal}
\ee
with $T_0$ is today's temperature.

From \eq{rp}, one can verify that the following class of Lagrangians $U$ 
\be
U=\lambda_y\;Y^{\frac{1+w}{w}}+\lambda_b\;b^{1+w}, 
\label{wmodel}
\ee
with $\lambda_b$ and $\lambda_y$ constants, leads to a constant barotropic 
equation of state
\be
p=w\;\rho,\;{\rm with}\;\;w'=0\to c_s^2=w \, , \qquad {  w
  \neq -1}  \, .
\ee
The same is true for
\be
U= b^{1+w}\;f\left(\frac{Y}{b^w}\right) \, ;
\label{wmodel1}
\ee
where $f$ is a generic function of its argument.
A list of  concrete interesting examples suitable to describe
important eras of our Universe is given bellow.
 \begin{itemize}
\item Radiation domination era $(w=1/3)$ 
$:
U=\lambda_y\;Y^4+\lambda_b\; b^{4/3}  
$ or $ U= b^{4/3} \;f(Y\;b^{-1/3}) $.
\item Matter domination ($w=0)$: $U=\lambda_b\;b$ or $U= b\;f(Y)$.
\item
Cosmological constant $(w=-1)$:
$U=f(b\;Y)$.
\end{itemize}

\subsection{Perturbed FLRW Universe} 
In this section we will show as the previous non perturbative equations \eq{dpi} are implemented at perturbative level around a FRW space time background.
The scalar perturbations of  FLRW Universe in the Newtonian gauge are 
\be
ds^2=a^2 \left[ \left(-1+2 \, \Psi \right) dt^2+\left(1+ 2 \, \Phi
  \right) d\vec{x}^2 \right]\, .
\ee
{ In the presence of perfect fluids, no anisotropic stress
  contribution to the EMT is present and}  we can set $\Psi=\Phi$.
At linear perturbation level the   EMT tensor can be obtained
from \eq{emt} by performing the following substitutions 
\be
\rho\to\rho+\delta\rho,\quad p\to p+\delta p,\quad
u_0=a\;(1+\Phi),\quad
u_i={ a} \, \partial_i  v \, .
\ee
For scalar perturbations the linearised perturbed Einstein equations
are given by
\ba 
&&  a^2 \;\delta\rho=4\, \plm^2\,\left[k^2 \, \Phi +3\; \mathcal{H}\;( \mathcal{H} \, \Phi+\,\Phi ')\right] \, ;\label{poisson}\\ 
&& 3\;{\cal H}^2\;(1+w)\; v \,=2 \,({\cal H}\, \Phi+\Phi') ; \label{vel}\\
&& a^2\;\delta p=- 4\;\plm^2 \;\left(\Phi''+3\;{\cal H}\;\Phi'-3\;w\;{\cal H}^2 \;\Phi \right)\, .
\label{dynphi}
\ea
From the definition of eq \eq{dp} it is clear that the expansion of $
\pmb{ \Gamma}$ start from the first order and  we will denote $\pmb{\Gamma}_{\rm order \;one} =\Gamma$.
The key equation that dictates the dynamics of the gravitational
potential $\Psi$ can be derived by combining (\ref{poisson}),
(\ref{dynphi}) and  the expansion of (\ref{dp})  
\be\label{mee}
\Phi'' +3\; \mathcal{H} \;\left(\mathit{c}_{\mathit{s}}^2+1\right) \Phi '+
   \left[k^2 \;\mathit{c}_{\mathit{s}}^2+3 \;\mathcal{H}^2\;
   \left(\mathit{c}_{\mathit{s}}^2-w\right)\right]\;\Phi +
 \frac{a^2}{4 \, \plm^2} \, \Gamma^{}=0 \, .
\ee
{  We stress that $\Gamma$ is the first order part of equation} \eq{dpi}.
Of course, to solve  (\ref{mee}) we need to know $\Gamma^{ }$ and, as we will
see, the effective field theory approach is the perfect tool
for that.

A particular important quantity to set the initial conditions for
cosmological perturbation is the comoving  curvature
perturbation~\cite{Liddle:1993fq,Lyth:1998xn}
\footnote{It can be shown that the ${\cal R}$ variable determines the
  curvature of equal time hypersurfaces in the comoving gauge.}
\be\label{defR}
{\cal R}\equiv -\Phi-{\cal H}\;v=
-\Phi-\frac{2\, (\Phi'+{\cal H}\, \Phi)}{3\;{\cal H}\;(1+w)}
\, ;
\ee
which satisfies the following first order differential equation
equivalent to \eq{mee}
\be\label{eqR1}
6\;(1+w)\;{\cal H}\;{\cal
  R}'=4\;k^2\;c_s^2\;\Phi+\frac{a^2}{\plm^2}\,\Gamma \, .
\ee
Similarly,  we can use also the curvature of uniform-density
hypersurface $\zeta$~\cite{Bardeen:1983qw} defined by
\be\label{defZ}
\zeta =-\Phi- \frac{\delta \rho}{3\;(\rho+p)} =-\frac{1}{3\;(1+w)}\;\left[\left(
5+3\;w+\frac{2\;k^2}{3\;{\cal H}^2}
\right)\;\Phi+\frac{2\;\Phi'}{\cal H}\right] \, ;
\ee
where the last equality follows from \eq{poisson}. By using again
\eq{mee}, the scalar $\zeta$ satisfies 
\be\label{eqZ1}
6\;(1+w)\;{\cal
  H}\;{\zeta}'-2\;k^2\;(1+w)\;\zeta=\frac{a^2}{\plm^2}\,\Gamma+
\frac{2\;k^2}{9\;{\cal H}^2}\;\left[2\;k^2+9\;(1+w)\;{\cal H}^2
\right] \;\Phi \, .
\ee
The  difference between   ${\cal R}$  and $\zeta$ is
given by 
\be 
\label{eqZRk}
\zeta-{\cal R}=-\frac{2\;k^2\;\Phi}{9\;(1+w)\;{\cal
    H}^2} \,.
\ee
By using \eq{eqR1} to eliminate $\Phi$,  we can write 
\bea
\label{eqZR}
\zeta-{\cal R} &=&-\frac{{\cal R}'}{3\;c_s^2\;{\cal H}}+
\frac{a^2\;\Gamma}{18\;\plm^2\;c_s^2\;(1+w)\;{\cal H}^2} \, ;
\ea
where we used   \eq{eqZ1} to derive the last relation. 
Equation \eq{eqZR} makes evident that  $\zeta$ and ${\cal R}$ differ
when 
\begin{itemize}
\item
entropic perturbations are present, namely when $\Gamma \neq 0$;
\item for adiabatic perturbations,
${\cal R}$ has  an increasing mode, that is when ${\cal R}$
 grows with the scale factor $a$. As an example, taking $c_s^2$
  constant and ${\cal R}\sim a^r$,  if $\Gamma=0$, we have 
  that  $\zeta-{\cal R} \propto{\cal R}'/{\cal
    H}\sim a^r$ and increases when $r>0$.
\end{itemize}
From EMT
conservation~\cite{Wands:2000dp} in the limit $k \to 0$ (large scale), it
follows that the only source of non-conservation of $\zeta$ for
superhorizon modes is
precisely due to the non adiabatic part of the pressure variation, namely
\be
\zeta'  = \frac{{\cal H} \; \Gamma^{ }}{(\rho+p)} = \frac{a^2 \; \Gamma^{ }}{6\;(1+w)\;{\cal H}\,{ \plm^2}} \, ;
\ee
thus, by integrating in redshift space
\be
\zeta =   \int_0^a  \frac{\bar a\;\Gamma^{ }(\bar a)}{6\;(1+w(\bar a)) \;{\cal H}^2(\bar a)} \, d\bar a \, .
\label{zetaeq}
\ee
While adiabatic initial  conditions are specified giving  $\zeta$ or
equivalently ${\cal R}$, at early time deep in radiation domination,
isocurvature initial conditions correspond to formally setting
\be
\lim_{a\to 0} \zeta =0 ,\qquad \lim_{a\to 0} {\cal R} =0  .
\ee
Clearly to have a closed set of evolution equations
we need to know $\Gamma$. For instance,  in eq \eq{mee} we expect that
generically $\Gamma(\Phi,\,v)$ and its structure is dictated by the equation
of state of our fluid. 
In presence of a perfect fluid the $\Gamma$ structure is dictated by a single combination of fields whose form will be the subject of the next section.

 \subsection{Cosmological Perturbations and fluid EFT}
The EFT description of a perfect fluid can be used to match the
standard cosmological perturbation in order to determine 
$\Gamma^{}$ entering in (\ref{mee}). The St\"uckelberg
scalars can be expanded as
\be
\Phi^0\equiv\phi(t)+\pi_0,\qquad \Phi^a\equiv x^a+V^a+\partial^a\pi_L
\ee
with $\partial_aV^a=0$. In the scalar sector 
only the scalar perturbations $\pi_L$ and $\pi_0$ are relevant. The
conservation of the EMT (\ref{emt}) is equivalent to the scalars
equation  of motion; in particular at the background level in FLRW we
have that~\cite{usthermo,classus}
\be
\phi'' + {\cal H} \, ( 3 \, c_b^2 -1) \, \phi' =0 \, ;
\label{phidd}
\ee
%
%
with $c_b^2$ given in \eq{cb}. When $c_b$ is constant we get that
\be
\phi'= \varphi_0 \; a^{1- 3 \, c_b^2} \, , \qquad \varphi_0 =
\text{constant} \, .
\ee
The expansion of the basic operators of the EFT in the Fourier basis  at the linear order
reads 
\ba
  b=\frac{1}{a^3}\;\left(1- k^2\,\pi_L+3 \,  \Psi  \right),
 \, \; Y=\frac{\phi'}{a}\left(1+\,\Psi+\frac{\pi_0'}{\phi'} \right)   , \, \; \label{Ybex}
 u^0=\frac{1}{a}\left(1+\Psi\right),\quad 
 u^m=\frac{i \, k^m  v}{a} \, .
\ea
Cosmological perturbations for a generic medium described by a
scalar effective theory can be found in~\cite{classus}. For the
benefit of the reader we rederive the basic relations in the case of a
perfect fluid. By using (\ref{rp}), (\ref{ch1}) and  (\ref{Ybex}),  
the perturbed  hydrodynamical variables can be rewritten in terms of 
$\pi_L,\,\pi_0$ and the gravitation potential $\Psi$ as follows
\ba
&&\delta \rho=-\frac{6 \, \plm^2}{a^2}\;(1+w)\;{\cal H}^2\;(3\;\Phi+k^2\;\pi_L)+\frac{\phi'}{a^4}\;\delta\sigma
\\
&&\delta p=-\frac{6 \, \plm^2}{a^2}\;c_s^2\;(1+w)\;{\cal H}^2\;(3\;\Phi+k^2\;\pi_L)+\frac{c_b^2\;\phi'}{a^4}\;\delta\sigma
\\
&& v=-\,\pi_L'   
\\
&&\delta \sigma= \frac{2\,M_0 \, \plm^2}{\phi'}\;\left[ (\Phi+c_b^2
  \left(3\;\Phi+k^2\;\pi_L) \right)
+\frac{\pi_0'}{\phi'}\right]
\ea
where 
$
M_0=\frac{a^2\;\phi'}{2 \, \plm^2}\;U_{YY} \, .
\;$
  The expansion of  (\ref{dp}) at the linear level allows to find the key relation  that gives
$\Gamma$ as a function of $\delta \rho$ and the entropy per particle
perturbation $\delta \sigma$ 
\ba
\delta p=c_s^2\; \delta \rho  +\frac{\phi '\;
 \left(c_b^2-c_s^2\right)}{a^4}\;\delta \sigma   \;\;
\Rightarrow  \;\;\Gamma =\frac{\phi '\;
 \left(c_b^2-c_s^2\right)}{a^4}\;\delta \sigma \, ;
\ea
where $c_s^2$ and $c_b^2$ are given by \eq{cs} and \eq{cb}, evaluated at
the zero order on the FRLW background \eq{FRW}. Actually, the very
same relation holds for a generic medium, see~\cite{classus}. The fact that for a perfect fluid  $\dot \sigma=0$, implies
that $\delta \sigma$ is  time independent ($\delta \sigma'=0$), or
equivalently  in Fourier space
\be
\delta \sigma(k,\,t)=\delta \sigma_0(k) \, .
\ee
As a result,  $\Gamma $ can be  factorised in a part that depends on $t$
and in a part $\delta \sigma_0$ that depends on the comoving momentum only
\be
 \Gamma(t,k) \equiv \underbrace{\frac{\phi '\;
   \left(c_b^2 -c_s^2 \right)}{a^4 }}_{\text{$t$-dependent}}\;\underbrace{\delta
 \sigma_0}_{\text{$k$-dependent}} \, .
\label{split}
\ee
The time dependence  of $\Gamma$ is so defined uniquely by the evolution of the background scale factor $a(t)$ and the thermodynamical quantities $c_s^2(t)$ and $c_b^2(t)$ (needed also to set the time dependence of $\phi'$ \eq{phidd}) always computed at background level \eq{cs}, \eq{cb}.
The above relation turns \eq{mee} in a closed equation for the
gravitation potential $\Phi$ with $\delta \sigma_0$ playing the role
of a source term. 
\subsection{Dynamics of ${\cal R}$ and $\zeta$ and the Weinberg theorem}
By definition ${\cal R}$ depends on $\Phi$ and
$\Phi'$ and the fact that $\delta \sigma_0$ is time independent allows
to write a closed second order differential equation for ${\cal R}$. Indeed, from  (\ref{defR}) and  (\ref{eqR1}) it
follows that
\be
\begin{split}
& {\cal R}''+\left[(2+3\;w-3\;c_s^2)\;{\cal H}- \frac{
     2 \,  c_s'}{c_s^2} \right]{\cal R}' +
k^2\;c_s^2\;{\cal R}+f(t) \,   \delta\sigma_0=0 \, ;\\[.2cm]
& f(t) = \frac{\phi' \, \left \{3 \,  {\cal H}\left(c_b^2-c_s^2\right) \left(2
      \, c_b^2-w-1\right) +2 \,c_b^2 \left[
      \log( c_s^2/c_b^2)\right]' \right \}}{12 \, \plm^2 \, a^2 \,
    (1+w) \, {\cal H}} \, \, .
\end{split}
\label{eqR2pf}
\ee
Notice that the above equation can be rewritten  also as
\be
\left[\frac{a^2 \;(1+w)}{  \, c_s^2}\right]^{-1} \frac{d}{dt}
\left[\frac{a^2 \;(1+w)}{ \, c_s^2} \, {\cal R}' \right] +
k^2\;c_s^2\;{\cal R}+f(t) \,   \delta\sigma_0=0 \, .
\label{eqR2pfn}
\ee
Adiabatic modes  for perfect fluids are characterised by the
global choice $\delta \sigma_0=0$, as discussed in detail
in~\cite{usthermo,classus}; for super horizon scales,  characterised by $k^2\ll (1+w)\;{\cal H}^2$ and $\delta
\sigma_0=0$, the dynamics of ${\cal R}$  can be easily read off from (\ref{eqR2pfn})  
\be
{\cal R} = {\cal R}_0 +   {\cal R}_1 \, \int^t dt' \,  \frac{c_s^2}{a^2 (w+1) }
 \qquad \qquad k^2\ll (1+w)\;{\cal H}^2 \, ;
\label{shRsol}
\ee
where $ {\cal R}_{0,1}$ are integration constants. Thus in general,
even for perfect fluids, it is not true that adiabatic
super horizon modes are constant, indeed from \eq{pG}
\be
\frac{d{\cal R}}{da} = \, \frac{{\cal R}' }{a {\cal H}}=  
 {\cal R}_1\; 
\frac{\left[\;{ 3\; w\; (1+w) \;\mathcal{H}-w'}\;\right]}{3 \;a^3\; (1+w)^2 \;\mathcal{H}^2}
 \qquad \qquad k^2\ll (1+w)\;{\cal H}^2 \, .
\label{Ra}
\ee
 Whenever, for large $a$, ${\cal R}$ is not
dominated by ${\cal R}_0$, the Weinberg theorem is violated; that
happens if the integral in (\ref{shRsol}) is a growing function of
$a$. 
\\ A sufficient condition for the violation of the Theorem is that  
\be
\frac{d{\cal R}}{da} \sim a^\beta \, , \quad \beta \geq -1
\quad \text{for large } a \, .
\label{wvcond}
\ee
When (\ref{wvcond}) holds, the super horizon constant mode ${\cal R }
= {\cal R }_0$ becomes sub-leading with respect to the growing mode
proportional to  ${\cal R }_1$. On the other  hand,  when $\beta <-1$ the mode
proportional to  ${\cal R }_1$ becomes sub-leading and ${\cal R }$ is
conserved. The bottom line is that adiabaticity is not sufficient to
guarantee that  ${\cal R }$ is conserved  for super horizon
perturbations. 
Notice that taking the limit $k \to 0$ in (\ref{eqR1}) is not sufficient to
guarantee the conservation of ${\cal R }$ on superhorizon scales;  it  simply shows
that there is a mode with such property; however ${\cal R}$ satisfies
a second order equation differential equation and one has to check what
happen to all independent modes on superhorizon scales and thus (\ref{eqR2pf})
is needed. 

As as an example take the following parametrization of the equation of
state
\be
 w = -1+a^\eta \, ;
\label{wpar}
\ee
where $\eta$ is a constant. From the definition of $c_s^2$ it follows
that
\be
c_s^2 = a^\eta-1 -\frac{\eta}{3} \, .
\ee
Then, for adiabatic super horizon perturbations, we have that 
\be
\frac{d {\cal R}}{da} ={\cal R}_1 \, \frac{3 \, a^{-3} -(\eta+3)
  a^{-\eta-3}}{3 \, {\cal H}} 
\label{Rdpar}
\ee
Notice that for large $a$, $w \to -1$ if $\eta <0$ and the metric is
close to dS~\footnote{We denote the curvature scale of dS by $\bar H$.},
e.g.  $a \sim - \frac{1}{ \bar H t}$ and it is suitable to describe an
inflationary phase of the Universe. In this case ${\cal H} \approx - a
\, \bar{H}$ and from (\ref{Rdpar})
\be
\frac{d {\cal R}}{da} ={\cal R}_1 \,  \frac{(\eta+3)
  a^{-\eta-4}-3 \, a^{-4}}{3 \bar H} \, .
\ee
Thus, the Weinberg theorem is violated when 
$\eta<-3$ see \cite{Kinney:2005vj,Namjoo:2012aa,Motohashi:2014ppa,Akhshik:2015nfa,Akhshik:2015rwa}. \\
Finally, in the case of constant sound speed, namely $c_s^2=w=$constant,  we get { (after a suitable redefinition of the integration constants)}
\be
{\cal R} ={  \bar{{\cal R}}_0 +   \bar{{\cal R}}_1} \, a^{3(w-1)/2}  \qquad \qquad
k^2\ll (1+w)\;{\cal H}^2 \, .
\ee
As expected,  ${\cal R}  \approx  { \bar{\cal R}}_0$ when $w \leq 1$.

Thanks to the relation \eq{eqZR}, once the dynamics of ${\cal R}$ is given,
$\zeta$ is completely fixed  by using (\ref{split}) 
\be
\begin{split}
\zeta-{\cal R}
&= -\frac{{\cal R}_1}{3\;a^2(w+1) {\cal H}} +
\frac{\phi' \, (c_b^2- c_s^2) \, \delta \sigma_0}{ { 18} \;\plm^2\, a^2 \, c_s^2 \,
  (1+w)\,  {\cal H}^2 }  \, .
\end{split}
\label{zetaRrel}
\ee
Of course, proceeding as for ${\cal R}$, one can determine a closed
evolution equation for $\zeta$,  however it is equivalent to
(\ref{eqR2pf}) and (\ref{zetaRrel}). For adiabatic perturbation, for
which $\delta \sigma_0=0$, we can work out the following simplified cases:
\begin{itemize}
\item When $w$ is constant we get that
\be
\zeta-{\cal R}= \bar{ {\cal R}}_1 \; a^{\frac{3\;(w-1)}{2} } 
\,.
\ee
so, for barotropic fluids with constant equation of state with $w \leq1$,
super horizon perturbations do not distinguish ${\cal R}$ from $\zeta$.

\item This is not necessary the case in a inflationary phase driven by a perfect fluid with
adiabatic perturbation. By using the parametrisation
(\ref{wpar}) with $\eta <0$  
we have for adiabatic super horizon perturbations
\be
 {\cal R} = {\cal R}_0 + {\cal R}_1 \frac{\left({ 1-a^{-\eta} }\right)}{{ 3\;\bar H}\;a^3} \, ; \qquad  \zeta = {\cal R}_0+ \frac{{\cal R}_1}{3\;\bar H\; a^3 } \, .
\ee
Thus, though ${\cal R}$ can violate  the Weinberg Theorem for $\eta<-3$, $\zeta$ is
always conserved for large $a$ \cite{Wands:2000dp}. 
\end{itemize}

Going back to entropic perturbations,
when $c_b^2$ and $w$ are constant ($c_s^2=w$), from
(\ref{Ydot})  one can   find   $\phi'=\varphi_0\;a^{1-3\;c_b^2}$ and determine the
time dependent part of $\Gamma $ 
\be
\Gamma  =\varphi_0\;a^{-3\,(c^2_b+1)}
   \left(c_b^2-c_s^2\right)\;\delta \sigma_0  \, ;
\ee
which can be used in (\ref{zetaeq}) to determine the  superhorizon
entropic contribution to ${\cal R}$ or $\zeta$  given by
\ba
{\cal
  R}={\cal R}_0 -\frac{ \varphi_0\;    \delta\sigma_0(k)}{18\;(1+w)\;{\cal
    H}^2_0\;\plm^2}\;a^{3\;(w-c_b^2)} \,
\ea
when (\ref{wvcond}) is not satisfied and the Weinberg theorem holds. 
Note that in order to implement the initial condition ${\cal R}\to 0$ for $a\to 0$ we need 
 $w-c_b^2\geq 0$  at early time.

\section{Multifluids}
\label{section:multi}
The case of a collection of perfect fluids which interact only
gravitationally  is similar to the single fluid treatment  \cite{Malik:2002jb}. Each component has an
individually conserved EMT of the form (\ref{emt}) with energy density
and pressure $\rho_i$, $p_i$, ``equation of state'' $w_i=p_i/\rho_i$
and adiabatic sound speed $c_{s i}^2= p_i'/\rho_i'$. For a multifluid
system is also convenient to define 
 \ba
 &&\rho=\sum_i\;\rho_i,\qquad p=\sum_i\;p_i,
 \qquad
 \omega_i\equiv \frac{\rho_i}{\rho}\\
 && w=\frac{p}{\rho}=\sum_i \, \omega_i\;w_i,\quad
 (1+w)\;c_s^2=\sum_i(1+w_i)\;\omega_i \, c_{s_i}^2 \, .
 \ea
The definition of $c_s^2$ is such that $p' = c_s^2 \, \rho'$.
The perturbed Einstein equations (\ref{poisson}) and (\ref{vel}) can be used to determine directly the
perturbed  total energy density and velocity defined by
\be\label{dtot}
\delta \rho=\sum_i\;\delta\rho_i \, ,  \qquad 
(\rho+p)\,  v_{\text{tot}}=\sum_i\;(1+w_i)\;\rho_i \,  v_i \, , \qquad
\delta = \frac{\delta \rho}{\rho} \, .
\ee
From the generalisation of (\ref{dp}) to the  multifluid case we have 
\be
dp_i=c_{s_i}^2\;  d\rho_i+ { \Gamma}_i  \;\; \Rightarrow \;\; \delta p_i 
=c_{s\,i}^2\;   \delta \rho_i + \Gamma_i^{ } 
\ee
Besides $\delta \rho$, $v_{\text{tot}}$ given by
\eq{dtot}, the total pressure variation can be written as 
\be\label{totalG}
\delta p = \sum_i  \;\delta p_i = \sum_i \;c_{s_i}^2 \;  \delta
\rho_i +  \Gamma^{ }_{\text{int}} \equiv  c_s^2 \, \delta \rho + \Gamma^{}_\text{tot} \, , \qquad
 \Gamma^{ }_\text{tot}= \Gamma^{ }_{\text{int}} +
\Gamma^{ }_{\text{rel}}  \, ,
\ee
where
\be
\Gamma^{ }_{\text{int}} = \sum_i \;\Gamma_i^{ } \, , \qquad \quad \Gamma^{ }_{\text{rel}} = \rho \; \sum_{i \neq
  j} \;\frac{(c_{s_i}^2-c_{s_j}^2)\;(1+w_i)\;(1+w_j) \; \omega_i \; \omega_j
}{(1+w)}\; S_{ij}  \, ,
\ee
and energy density and velocity perturbations of each component have
been conveniently  parametrised  in terms of 
\ba
&& S_{i j}=-3\;{\cal H}\;\left(\frac{\delta
     \rho_i}{\rho_i'}-\frac{\delta \rho_j}{\rho_j'}\right) =
 \frac{\delta_i}{(1+w_i)}-\frac{\delta_j}{(1+w_j)}  \qquad  \qquad
 \qquad \, i
 \neq j  \, ;\\
&& v_{ij} = v_i - v_j  \, , \qquad \Gamma_{ij}^{ } =
\frac{ \Gamma_i^{ }}{(1+w_i)\;\rho_i } -
\frac{\Gamma_j^{ }}{(1+w_j) \;\rho_j}
  \qquad \qquad \qquad i
 \neq j \, .
 \ea
We will be mostly interested to the case of two fluids and  we set
$S_{12}=S$, $v_{12} = v_d$ and $\Gamma_d= \Gamma_{12}^{ }$. 
The gravitational potential satisfies the equation
\ba\label{master}
&& \Phi '' +3 \mathcal{H}
   \left(c_s^2+1\right) \Phi ' +\Phi  \left[k^2 \, c_s^2 + 3
     \left(c_s^2-w\right) 
   \mathcal{H}^2\right] 
{ +}\frac{a^2\;\Gamma_{\text{tot}}}{4 \, \plm^2} =0 \, ; \\[.2cm]
&&   \Gamma_{\text{tot}} = \Gamma_1+\Gamma_2 + \frac{\rho _1\; \rho _2\; \left(w_1+1\right) \left(w_2+1\right)
   \left(c_{{s_1}}^2-c_{{s_2}}^2\right)}{\rho _1
   \left(w_1+1\right)+\rho _2 \left(w_2+1\right)} \;S  \, ; 
   \ea
 while
\ba
&&S' = k^2 \,  v_d - 3 \, {\cal H} \,  \Gamma_d \, ; \label{Sp}\\[.2cm]
&& {v'}_d=\frac{4 \,\plm^2 \,k^2 \,(c_{s_1}^2 -c_{s_2}^2)}{a^2 \,(1+w)\, \rho}\,
\Phi- \frac{(3\, c_{s_2}^2 -1)(1+w_1) 
\, \omega_1 +(3\,
  c_{s_1}^2 -1)(1+w_2) \, \omega_2}{(1+w)} \,{\cal H} \; v_r  \nb \\
&& \qquad + \frac{c_{s_2}^2(1+w_1) \, \omega_1 + c_{s_1}^2 (1+w_2) \, \omega_2}{1+w} \, S -\Gamma_d\, .
\ea
It's easy to see that, in the limit $k\to 0$,  $ v_d$  
decouple from the evolution of the other variables.

An important gauge invariant quantity that controls  linear structure formation
can be obtained from the total matter contrast $\delta$ by defining
\be
{ 
\Delta = \delta{ -} 3 \, (w+1) \, \mathcal{H} \, 
  v_{\text{tot}} }  \, ;
\ee
the Hamiltonian constraints  \eq{poisson} reads
\be
{ 
3 \, \mathcal{H}^2  \, \Delta  { -}  2 \, k^2 \, \Phi =0 }\, .
\label{hamdelta}
\ee
From (\ref{hamdelta}) and (\ref{master}) we can write the following evolution
equation for the gauge invariant matter contrast $\Delta$
\be
\frac{d^2 \Delta}{da^2} +\frac{3 \left(2 \, c_s^2-5 \, w+1\right)}{2 a}\;  \frac{d \Delta}{da} +
\frac{\left[2 \, c_s^2 \left(\frac{k^2}{\mathcal{H}^2}+9\right)+3 \, w \,(3
   \, w-8)-3\right]}{2 a^2}\;\Delta { +}\frac{k^2 \,\Gamma_{\text{tot}}}{6 \, \plm^2 \, \mathcal{H}^4} =0 \, ;
\label{deltaeq}
\ee
where we have considered $\Delta$ as a function of $a$.
As for $\Phi$, $\Delta$ is sourced by both the intrinsic and relative
entropy perturbations. The evolution  of $\Delta$ is a popular test
from deviation from $\Lambda$CDM.\\
In the multifluid case ${\cal R}$  is defined as
\be\label{defRm}
{\cal R} = -\Phi - {\cal H} \, v_{\text{tot}} =-\Phi - \frac{2(\Phi'+ {\cal
    H} \, \Phi)}{3 \,  {\cal H}(1+w)} \, .
\ee
As in the case of single fluid,
${\cal R}$ satisfies a conservation equation that is basically the
same as (\ref{eqR1}) with the replacement $\Gamma \to \Gamma_{\text{tot}}$ 
and from our effective description 
\be
\Gamma_i = \frac{\phi_i'\;(c_{b_i}^2 -c_{s_i}^2)}{a^4} \, \delta \sigma_i
\, , \qquad i=1,2 \, .
\label{Gamma}
\ee
Proceeding as for the single fluid case, we get the following evolution
equation for ${\cal R}$
\be
\mathcal{R}''+
   \left[\mathcal{H} \left(3\, 
   w+2 -3 \, c_s^2\right) -\frac{c_s^2{}'}{c_s^2}\right]\mathcal{R}' +k^2
\,  c_s^2\, \mathcal{R}+\frac{a^2 \, \Gamma_{\text{tot}} \left[2 \,
    c_s^2{}'-3 \, c_s^2 \, (w+3) \mathcal{H}\right]}{12 \, \plm^2  \, (w+1) \, \mathcal{H}
   c_s^2}-\frac{a^2 \Gamma_\text{tot}'}{6 \, \plm^2 (w+1)
   \mathcal{H}}=0 \, .
\ee
From (\ref{Sp}), for superhorizon modes, $S'=- 3 \, {\cal H} \, \Gamma_d$;
thus for adiabatic and superhorizon perturbations: $\delta
\sigma_{1/2}=S=0$ and $\Gamma_{\text{tot}} =0$. As a result, 
the very same consideration for a single fluid case applies and
the Weinberg theorem can be violated when (\ref{wvcond}) holds.
Similarly $\zeta$ is still defined by (\ref{defZ}) with $p$ and $\rho$
and $\delta \rho$ refer to the sum of the various fluid components.
When the Weinberg theorem holds we have that,  likewise the single fluid case, 
\be
\zeta' = {\cal R}' = \frac{ {\cal H} \;
  \Gamma^{ }_{\text{tot}}}{(\rho+p)} \Rightarrow  \zeta =  {\cal R}=
  \int_0^a  \frac{\Gamma^{ }_{\text{tot}}}{(\rho+p)\; a'} \, da' \, .
\label{zetaeqm}
\ee
The bottom line is that even for  multi-component perfect fluids the Weinberg theorem
in general does not hold; and
even when it holds, if fluids are non-barotropic, ${\cal R}$ and
$\zeta$ are not conserved due to entropic effects. 
Entropic perturbations can have two different origin:  \underline{intrinsic}
fluctuation of the entropy per particle for each fluid component due
to its non-barotropic nature and \underline{relative} coming from non-adiabatic
variation of $p$ due to the relative difference
of density contrast of the various components. The EFT approach to
perfect fluids gives the complete form of
$\Gamma_i^{ }$~\footnote{Actually (\ref{dpi}) gives a 
fully non-perturbative
  expression.}
and with the help of (\ref{Gamma}) we have a closed variational system of equations for
cosmological perturbations.

 \section{Universe evolution and entropic perturbations}
 In what follows  we will study the effect of entropic perturbations
 in various fluid models. The benchmark model is of course  
  $\L$CDM for which  the entropy source is the presence of
  uncoupled barotropic fluids with a non-trivial relative pressure perturbation $\Gamma_{rel}$. 
The very  same  background evolution and entropic perturbations of $\L$CDM can
be obtained by using a perfect single-fluid model  described by a Lagrangian of the
form $U(b,\,Y)$, see appendix \ref{section:ONE-baro}. In such a case  the origin of entropic perturbations
is the intrinsic $\Gamma_{int}$ pressure  perturbation.
We  study a more physical multi-fluid system composed by a barotropic radiation component (photons and neutrinos)
and a generic dark fluid (DF), described  by a Lagrangian  of the form $U(b,\,Y)$,
representing the dark matter (DM) and dark energy (DE) system.
An analytical analysis is carried out only for superhorizon scales for
simplicity.

\begin{table}[h!]
  \footnotesize
  \small
  \renewcommand\arraystretch{1.3}
  \centering
  \begin{tabular}{|| l |c|c|c| | ||}
    \hline
$\Lambda$CDM  & $U(b,\,Y)$ & Radiation$\;+\;U(b,\,Y)$ \\
    \hline \hline
$\Gamma_{rel} $ &$\Gamma_{int} $& $\Gamma_{int} +\Gamma_{rel} $\\ \hline\hline
\end{tabular}
  \caption{Sources of Entropic Perturbations in the $\Lambda$CDM model, the single fluid described by a potential $U(b,\,Y)$ and the multi fluid model composed by a non interacting  barotropic radiation fluid and an entropic fluid $U(b,\,Y)$.}
  \label{tab:Leg}
 \end{table}

\subsection{The Benchmark model: $\Lambda$CDM}
\label{sect:LCDM}
In the concordance $\Lambda$CDM cosmological model  various barotropic
fluids contribute to the EMT. 
For simplicity we will consider here three perfect fluids: radiation (photons)
with $w_1=c_{s_1}^2=1/3$, DM with $w_2=c_{s_2}^2=0$ and 
DE with $w_{\L}=-1$. The more recent estimation of the cosmological
parameters can be found in~\cite{Planck-2015-par}. Being each  component 
barotropic  we have that $\Gamma_i=0$. In  $\Lambda$CDM  DE is just  a cosmological
constant, thus  such a system  is effectively a two-fluid model of DM and photons. In particular  the energy density $\bar\rho$, pressure $\bar p$, effective equation of state $\bar w$ and the sound speed $\bar{c}_s^2$ are
\ba
     &&\bar\rho
     = 6 \, \plm^2 \, H_0^2 \,
     \left(\Omega_{\L}+\frac{\Omega_{m}}{a^3}+\frac{\Omega_{r}}{a^4} \right),\qquad
     \bar p
     = 6 \, \plm^2 \, H_0^2 \,
     \left(\frac{\Omega_{r}}{3 \, a^4} -\Omega_{\L}\right),
     \\ && 
     \bar w =\frac{ p}{ \rho}=  \frac{
       (\Omega_{r}-3\;a^4\;\Omega_{\L})}{3\;(a\;\Omega_{m}+\Omega_{r}+
a^4\;\Omega_{\L})},\qquad
     \bar c_s^2  = \frac{p'}{\rho'}= \frac{ 4\;a_{eq}}
   { 3\;(4\;a_{eq}+3\;a )} \, ; \label{lcdm}
\ea
where $H_0$ is the  today Hubble parameter, $\Omega_{r}\simeq 10^{-4}, \; \Omega_{m}\simeq
0.25,\;\Omega_{\L}\simeq 0.75$ (with $\Omega_{r}+\Omega_{m}+\Omega_{\L}=1$), 
    $a_{eq}=\Omega_{r}/\Omega_{m}
    $ 
    and $a=1$ is the today's value.
For superhorizon modes    the  relative entropy
  perturbation between
DM and radiation $S=\delta_m- 4 \, \delta_r /3$
is conserved $S'=0$,  \eq{Sp}, with    solution    $S=s_0(k)$. The only contribution to the
non-adiabatic pressure variation comes from $S$ and we have
\be
 \Gamma_{\text{tot}} \equiv \Gamma_{\text{rel}}
 =\frac{8\, H_0^2 \, \plm^2 \, \Omega_{m}\, \Omega_{r}}{a^3\left(4 \,
     \Omega_{r}+3\, a\, \Omega_{m} \right)}\;s_0(k)
 \, .
\label{gammalcdm}
 \ee
The Weinberg theorem of course holds, indeed from (\ref{Ra}), we have that
for large $a$
\be
\frac{d {\cal R}}{da} =   \frac{4\; {\cal R}_1 \,\, \Omega_r \, \Omega_\L^{1/2}}{9 \,
  H_0 \, \Omega_m^2 }\;\frac{1}{a^2} +\text{O}\left(\frac{1}{a} \right)^3 \, .
\ee
The only source of non-conservation of ${\cal R}$ is due to the
contribution of non-adiabatic perturbation conserved at superhorizon
which, by using (\ref{zetaeqm}) and (\ref{gammalcdm}), leads to 
\be\label{SS}
 {\cal R}(a)= s_0(k) \;\frac{{  a}}{3\;a+4\;a_{eq}},\qquad \bar{{\cal R}}
 \approx \frac{1 }{3}\;s_0(k) \, ;
  \ee
where $ \bar{{\cal R}} = {\cal R}(a=1)$ is the present  value. The
standard choice of adiabatic initial conditions is equivalent to set
$s_0=0$. Planck~\cite{Planck-infl} limits roughly allow up to few
percents of non-adiabatic contribution from cold dark matter.   
     
     \section{Two-fluid Universe Models}
{  Let us  study a very simple modelling of our Universe in terms of two fluids:
radiation (photons)  and a second non-barotropic perfect fluid representing together 
dark matter and dark energy described by the Lagrangian $U(b,\,Y)$.
We neglect the effect of baryons.}
The dynamical equations in the multi-fluid case are described in
section \ref{section:multi}.
We use the subscript 1 to denote radiation which  is
characterized by
 \be
 c_{s_1}^2=w_1=\frac{1}{3} \, ,\qquad \rho_1=3\;p_1=6\, \plm^2 \,
 H_0^2\;\frac{\Omega_{r}}{a^4}, \quad \qquad \Gamma_{1}=0 \, .
 \ee
In particular, in the small $k$ limit (superhorizon), the dynamics of
${\cal R}$ can be deduced (see equations (\ref{Sp}) and
(\ref{zetaeqm})) from  the following  coupled set of
equations for $S$  ($S\equiv S_{12}$)  and ${\cal R}$
\be
\label{eqs}
\begin{split}    
 & S'=3 \, {\cal H} \, \frac{\phi'(c_b^2-c_{s_2}) \, \delta
   \sigma_0}{(1+w_2) \, \rho_2 \, a^4},\qquad
{\cal R}'=-\frac{ a^2\;\Gamma_{tot}}{6\;(1+w)\;{\cal H}} \
\, ;\\[.2cm] 
&
\Gamma_{tot}= \frac{\phi'(c_b^2-c_{s2}) \, \delta
  \sigma_0}{a^4}+{  6\,
\plm^2} \, \frac{S\;{\cal
    H}^2}{a^2\;(1+w)}\left(\frac{1}{3}-c_{s_2}^2\right)\;\frac{4}{3} \, (1+w_2)\;\omega_1\;\omega_2
\, ;    
\end{split}
\ee
where we have set $c_{b_2}^2 = c_b^2$ and $\phi_2' = \phi'$.
$S$ can be obtained by the integration of the first equation in
\eq{eqs}. Once $S$ is known, the second equation in \eq{eqs} can be
solved for ${\cal R}$.
 Two boundary initial conditions are needed to solve eqs \eq{eqs} and in order to select the entropic contributions we
 impose that 
\be
\lim_{a \to 0} {\cal R} =0 \, , \qquad \lim_{a \to 0} {\cal S} =s_0 \,.
\ee
Basically the spectrum of primordial non-adiabaticity is encoded in
 the  two initial conditions ($k$-dependent)   $s_0$ and
$\delta \sigma_0$.
In the following we will analyse some very simple models that can be easily solved  analytically.
One of the main request for the comparison in between them is the fact that the respective effective  equations of state   $w$ has to be marginally compatible with the 
$\Lambda$CDM one \eq{lcdm} over all the temporal range in between nucleosynthesis time  ($a=10^{-10}$) until the present time ($a=1$) with a reasonable   error  of less than 1\% \cite{ Planck-2015-par}.
 In synthesis the   models that follow are composed by radiation and a dark fluid that gives almost the same background evolution (starting from nucleosynthesis) as $\Lambda$CDM. The dark fluid is a single entropic perfect fluid whose perturbations are   analytically solvable on   superhorizon limit and with an equation of state
   as simple as possible to summarise all the above features.  
 
\subsection{Case 1}
\label{case1}
   As first example we take as dark component a  fluid with a Lagrangian $U$
composed of two terms which, when considered individually, would
describe non-relativistic matter representing DM with $w=0$, see (\ref{wmodel}),
and  a DE component, see (\ref{wmodel1}), with $w=-1$.
The Lagrangian $U$ is defined as
\be U=6 \, \plm^2 \, H_0^2\left[-\Omega_{
    m}\;b+ \Omega_{\Lambda}\;(b\, Y)^2
\right] \, .
 \ee
  Notice that having modelled the dark sector has a unique fluid we are
effectively considering an interacting DM and DE system.  
At the background level, see equations (\ref{Tscal}) and (\ref{phidd}) , we have 
 \be
\phi' = \varphi_0 \, a^{4} \,, \qquad c_b^2=-1 \, ,\quad
 c_{s_2}^2=0\, ;
 \ee
and the temperature of the dark fluid is given by $ T_{DF}=Y=\varphi_0  \, a^{3}$ \cite{Lima:2004wf}. 
The effective equation of state $w$ is exactly   the same of 
the one of $\Lambda$CDM and it is given by \eq{lcdm}
$
w =\bar w
$ 
 with $ \Omega_{ r}+\Omega_{ m}+ \Omega_{\L}=1$.\\
  The relative entropy for the dark sector/photons and the curvature
perturbation, given by \eq{eqs}, are
\bea
S = s_0 - \frac{ 4\;\Im_0\;a_{eq}  \;a^3 }{3\; } 
,\qquad
{\cal R}= \frac{s_0 \;a}{3\;a+4\;a_{eq}}-
\;\frac{4\;\Im_0\;a_{eq}\;a^4}{3\;(3\;a+4\;a_{eq})} \, ;
\ea
where we have defined
\be
\Im_0\equiv \frac{\varphi_0\;\delta \sigma_0}{8\;\plm^2\;H_0^2\;\Omega_r}
\ee
Note that the relative entropic contribution (proportional to $s_0$) is the same of  $\Lambda$CDM \eq{SS}.

\subsection{Case 2}
\label{case2}
 This time the dark sector Lagrangian  is  such that the
interaction between DM e and DE is different  
\be U=-6 \, \plm^2 \, H_0^2\; \left[ \frac{1}{3}(\Omega_m+\Omega_z)^{2/3}\;\left(\Omega_{
    m}\;\frac{b}{Y^2}+ \frac{\Omega_{z}}{(b\;Y)^2}\right)+\Omega_{\Lambda}
\right] \, .
 \ee
 Though the single terms describing DM and DE have the same equation
of state, $w=0$ and $w=-1$ respectively, the full equation of state  is different from the previous case. 
At the background level we have (see \ref{Tscal} and \ref{phidd})
 \be
\phi' = \varphi_0 \, a\;\sqrt{\k_1} \,, \qquad c_b^2=-1+\frac{\Omega_m}{\k_1^{3/2}} \, ,\quad
 c_{s_2}^2=\frac{2\;a^9\;\Omega_{z}}{\k_1^{3/2}} \, ;
  \qquad
 \ee
with $ \k_1\equiv(\Omega_m+\Omega_{z}\; a^{9})^{2/3} $ and $ \k_2\equiv(\Omega_m+ \Omega_{z})^{2/3} $.
The Dark Fluid has a temperature $  
 T_{DF}=Y=\varphi_0  \,\sqrt{\k_1} $; notice also the
   unusual asymptotics of $c_{s_2}^2$ for large $a$. The effective equation of state is
\be
w= \frac{3 \;a \;\kappa _2\; \Omega _m+4
\;   \kappa _1\; \Omega _r}{3\;
   \left(\kappa _1 \;\left(a^4\; \Omega
   _{\Lambda }+\Omega _r\right)+a\;
   \kappa _2\; \kappa
   _1^{3/2}\right)}-1 \, ;
\ee
 with $ \Omega_{ r}+\Omega_{ m}+ \Omega_{z}+ \Omega_{\L}=1$.
 For sufficiently small values of $\Omega_z$ it     reproduces the value  $\bar w$  of  $\Lambda$CDM.
From \eq{eqs} we have that 
\bea\label{SS2}
S = s_0 - \frac{  4\;\Im_0 \;\Omega_z \;a_{eq}}{3  \;\k_2} \;a^9,\qquad
{\cal R}= \frac{ \,a\, s_0 \, 
  \k_2}{4 \,a_{eq}\,
   \k_1+3 \,a\,
    \k_2}-
   \frac{4\,\Im_0\,
   a^{10} \,a_{eq}\,  \Omega_z}{3 
   \left(4\,   a_{eq}\,\k_1+3 \,a\,  
   \k_2\right)}
\ea
Notice that in the limit $ \Omega_m\gg \Omega_z$ (as it has to be imposed phenomenologically) we can approximate $\k_1=\k_2\sim \Omega_m^{2/3}$ and simplify considerably eqs  \eq{SS2}.
Again, in such a limit,   the relative entropic contribution (proportional to $s_0$) is the same of  $\Lambda$CDM \eq{SS}.

 \subsection{Case 3}
\label{case3}
 In this case we added to   the dark sector   an exotic component that alone would have
an equation of state $w_z$ and the DE component is just a cosmological
constant. We take 
\be U=6 \, \plm^2 \, H_0^2\; \left(-\Omega_{
    m}\;b+\;w_z\;\Omega_{z}\;Y^{\frac{1+w_z}{w_z}}-\Omega_{\Lambda}
\right) \, .
 \ee
 The interaction in the dark sector is among the two DM components.
At the background level we have
 \be
\phi' = \varphi_0 \, a^{1- 3 \, w_z} 
,\qquad c_b^2=w_z \, ,\quad
 c_{s_2}^2=\frac{w_z\,(1+w_z)\;\Omega_z}{\Omega_m\;a^{3\,w_z}+(1+w_z)\;
\Omega_z} \, ;
 \ee
and the temperature of the DF scale as $T_{DF}=Y= \varphi_0  \, a^{-3\, w_z} $.
The effective equation of state corresponds to
\be
w = \frac{a^{3\,  w_z} \left(\Omega _r-3 \, a^4 \, \Omega _{\Lambda
    }\right)+3 \, a
\,    w_z \, \Omega _z}{3 \, a^{3 \,w_z} \left(a^4 \, \Omega _{\Lambda
  }+a \, \Omega
   _m+\Omega _r\right)+3 \, a \,\Omega _z} \, ;
\ee
 with $ \Omega_{ r}+\Omega_{ m}+\Omega_{ z}+\Omega_{\L}=1$ and for $\Omega_z\to 0$ it goes  to $\bar w$.  From (\ref{Ra}) for large $a$ it follows that the Weinberg theorem holds
 as soon as $w_z> 1/6$
\be
\frac{d {\cal R}}{d a} = \frac{\Omega_\L^{1/2} \, \Omega_r}{9 \, H_0
  \left[\Omega_r +(1+w_z) \Omega_\L \right]^2 \, a^{6 \, w_z}} +
\text{O}\left(\frac{1}{a^{3 \, w_z -3}} \right)\, .
\ee
In the present case, the solution  of  \eq{eqs} is
\be
S = s_0 + \frac{3 \, \delta \sigma _0 \, \Omega _m \, a^{3 \, w_z}}{q(a)}
\, , \qquad {\cal R} =\begin{cases}  q(a)^{-1}\, \delta \sigma _0 \, \Omega _m \, a^{3
      w_z}+\frac{s_0}{3} & \frac{1}{6}<w_z
<\frac{1}{3} \\[.2cm]
q(a)^{-1}\, \delta \sigma _0 \, \Omega _m \, a^{3 \,w_z}&  w_z
>\frac{1}{3} \end{cases} \, ;
\label{eqR}
\ee
with 
\be
q(a) = 18 \, H_0^2\, \plm^2 \,
   \left(w_z+1\right) \, \Omega _z \, \left[\Omega _m \, a^{3
   w_z}+\left(w_z+1\right) \, \Omega _z\right] \, ;
\ee
and  $s_0=s_0(k)$ is a $k$-dependent integration constant. For
simplicity we have retained only the leading terms in $\Omega_r$, the
full expression for ${\cal R}$ is given in appendix \ref{appcase1}.
 The apparent singularity when $\Omega_z \to 0$ is not physical;
 indeed, in that  limit one should also send $\delta \sigma_0$ to
 zero, basically the dark sector becomes  a cosmological constant. 
A limit on $\Omega_{z}$ can be obtained from  primordial
  nucleosynthesis taking place roughly  at $a_n\sim 10^{-e_n}$ with
  $e_n\sim10$, imposing that  the expansion rate of the ${\cal
    H}(a_n)$ is close enough  to  the $\Lambda$CDM  one ;
  that is
\be
\label{OZ}
 \left. \frac{{\cal H}^2-{\cal H}_{\Lambda CDM}^2} {{\cal
       H}^2_{\Lambda CDM}}\right
 |_{a_n}\simeq\frac{\Omega_z}{\Omega_r} \;a_n^{1-3\;w_z}\leq
 10^{-2} \Rightarrow  w_z \leq \frac{e_n-2+\l}{3\;e_n}  \, ;
 \ee
where $\frac{\Omega_z}{\Omega_r} =10^{-\l}$. 
When $w_z <1/3$, ${\cal R} \to 0 $ for large $a$, while
when $w_z >1/3$ and $\delta \sigma_0 \neq 0$, ${\cal R} $ has a finite
limit showing how curvature  perturbations can be significantly
boosted by intrinsic entropic perturbations. \\
 Below we give the explicit results of  the special cases  $w_z=1$,
  corresponding kineton-like exotic component, and  also for $w_z=1/4$. 
  \begin{itemize}
  \item We
  have that
the case with $w_z=1$ (that we named case 3$a$) gives
 \ba\nonumber
  Y=b ,  \;w=\frac{-3 \;a^6\; \Omega _{\Lambda }+a^2\; \Omega _r+3
\;   \Omega _z}{3\; \left(a^6\; \Omega _{\Lambda }+a^3
 \;  \Omega _m+a^2 \;\Omega _r+\Omega _z\right)} ,
  \;c_{s_2}^2=\frac{2\;\Omega_z}{(2\;\Omega_z+a^3\;\Omega_m)} ,
  \;c_b^2=1  \, ;
%
\\ S=s_0+\frac{2\;\Im_0 \;a^3\;\Omega_m^2\;a_{eq}}{3 \;\Omega_z\;(2\;\Omega_z+a^3\;\Omega_m)},\qquad 
{\cal R}=
\frac{2\;a^3\; \Im_0\;  \Omega_m^2\;a_{eq} -4 \;a^2\;   s_0 \; \Omega_m\;\Omega_z\;a_{eq}}
{3\;\Omega_z \;\left( a^2\;(3\; a+4\;a_{eq} )\;\Omega_m+6\; \Omega_z
  \right)} \, .
\ea
\item
The case with $w_z=\frac{1}{4}$   (that we named case 3$b$)  gives 
\ba\nonumber
Y=b^{1/4},\; 
w=\frac{-12 \;a^4\; \Omega _{\Lambda }+3\;a^{1/4}\; \Omega _z+4\;
   \Omega _r}{12 \;\left(a^4\; \Omega _{\Lambda }+a\; 
   \Omega _m+  \;\Omega _r+a^{1/4}\;\Omega _z\right)},\;
   c_{s_2}^2=\frac{5\;\Omega_z}{4\;(5\;\Omega_z+4\;a^{3/4}\;\Omega_m)},\;
   c_b^2=1/4 \, ;
\\
 S=s_0+\frac{64\;\Im_0\;a^{3/4}\;\Omega_m^2\;a_{eq}}{15 \;\Omega_z\;(5\;\Omega_z+4\;a^{3/4}\;\Omega_m)},
\;\;
 {\cal R}=\frac{15\;s_0 \left(4\; a \;\Omega _m+5\;
   a^{1/4}\; \Omega _z\right)\; \Omega _z+64\;\Im_0\;a\;\Omega_m^2\;a_{eq}}
   {15 \; \Omega
     _z\;\left(4\;(3\;a+4\;a_{eq})\;\Omega_m+15\;a^{1/4}\;\Omega_z\right)
   } \, .
\ea
\end{itemize}
In Figure (1) and (2) we  show explicitly their time dependence.%

\subsection{Summary}
The features of the various cases considered are summarised in table
\ref{tab:Leg1}. Generically, the superhorizon
curvature perturbation can be written as
\be
{\cal R} = {\cal R}_{s_0} \, s_0 +{\cal R}_{\Im_0} \, \Im_0 \, .
\ee
In the table, for each case, $c_s^2$ and $c_b^2$ and  both
contributions $ {\cal R}_{{s_0}}$, ${\cal R}_{\Im_0}$ are shown.
\begin{table}[h!]
  \renewcommand\arraystretch{1.4}
  \centering
  \begin{tabular}{|c|c|c|c|c|}
 \hline\hline
 Universe  &  $c_{s_2}^2$     &  $c_b^2$ &  ${\cal R}_{{s_0}}$ & $ {\cal R}_{\Im_0}$ \\\hline
  \hline
    $\Lambda$CDM    &$ \frac{4\;a_{eq}}{3\;(3\;a+4\;a_{eq})}$& $  0$ &  $ \frac{a}{3\;a+4\;a_{eq}}$& 0\\
    \hline
  Case 1 
  &
    $ 0$ & $ -1$ &
    $\frac{a}{3\;a+4\;a_{eq}}$ & $-\frac{4\;a_{eq}\;a^4}{3\;(3\;a+4\;a_{eq})}$
    \\\hline 
  Case 2 
  &  $ \frac{2\;a^9\;\Omega_z}{\k_1^{3/2}}$ & $-1+ \frac{ \Omega_m}{\k_1^{3/2}} $&
    $\frac{2\;a\;\k_2 }    {3\;a\;\k_1 +4\;a_{eq}\;k_2 }$ & 
    $-\frac{4\;a_{eq}\;\Omega_z\;a^{10}}{4\;(3\;a\;\k_1+4\;a_{eq}\;\k_2)}$
    \\\hline
  Case 3$a$ 
     &
      $\frac{2\;\Omega_z}{2\;\Omega_z+a^3\;\Omega_m} $ &$1$
       &$-\frac{4\;a^2\;\Omega_m\;a_{eq}}{3\;( a^2\;(3\;a+4\;a_{eq})\;\Omega_m+6\;\Omega_z)}$&$\frac{2\;a^3\;\Omega_m^2\;a_{eq}}{3\;\Omega_z\;( a^2\;(3\;a+4\;a_{eq})\;\Omega_m+6\;\Omega_z)}$
    \\\hline
  Case 3$b$ 
     &$\frac{5\;\Omega_z}{4\;(5\;\Omega_z+4\;a^{3/4}\;\Omega_m)}$
    &$\frac{1}{4}$&$\frac{4\;a\;\Omega_m+5\;a^{1/4}\;\Omega_z}{4\;(3\;a+4\;a_{eq})\;\Omega_m+15\;a^{1/4}\;\Omega_z}$ &
     $\frac{64\;a\;\Omega_m^2\;a_{eq}}{15\;\Omega_z\;(4\;(3\;a+4\;a_{eq})\;\Omega_m+15\;a^{1/4}\;\Omega_z})$
      \\\hline
\end{tabular}
  \caption{Summary table for the cases: $\Lambda$CDM, case (1) with $U=
    6\,H_0^2\,(-\Omega_m\,b+ \Omega_\Lambda\;(b\;Y)^2)$, case (2) with
    $U=6\,H_0^2\,(\Omega_m\,\frac{b}{Y^2}+ \frac{\Omega_z}{(b\;Y)^2}-
    \Omega_\Lambda)$, case (3$a$) with $U=6\,H_0^2\,(-\Omega_m\,b+
    \Omega_z\;Y^2- \Omega_\Lambda)$, case (3$b$) with $U=6\,H_0^2\,(-\Omega_m\,b+ \frac{\Omega_z}{4}\;Y^5- \Omega_\Lambda)$.}
  \label{tab:Leg1}
 \end{table}
In figure \eq{Evol1} we  plotted the ratio ${\cal R}_{{\Im_0}}/\bar{\cal R}_{{\Im_0}}$ 
(with $\bar{\cal R}_{{\Im_0}}={\cal R}_{{\Im_0}}(a=1)$), while
in figure \eq{Evol2} the ratio ${\cal R}_{{s_0}}/\bar{\cal R}_{{s_0}}$ 
(with $\bar{\cal R}_{s_0}={\cal R}_{s_0}(a=1)$) for the
four cases of table \ref{tab:Leg1} as a function of $a$, having set $a=1$,
the today value of the scale factor. Notice that the above ratios are
independent from the normalisations $s_0$ and $\Im_0$ and highlight the time dependence of the various corrections. 
We have taken  $\Omega_m=0.25$, $\Omega_r=10^{-4}$;  $a_n$ is the
scale factor at nucleosynthesis $\sim 10^{-10}$ and finally
$a_{eq}=\Omega_r/\Omega_m$ corresponds to  matter radiation equality. 
 \begin{figure}[h!!]
\begin{center}
\includegraphics[width=13cm,height=7cm]{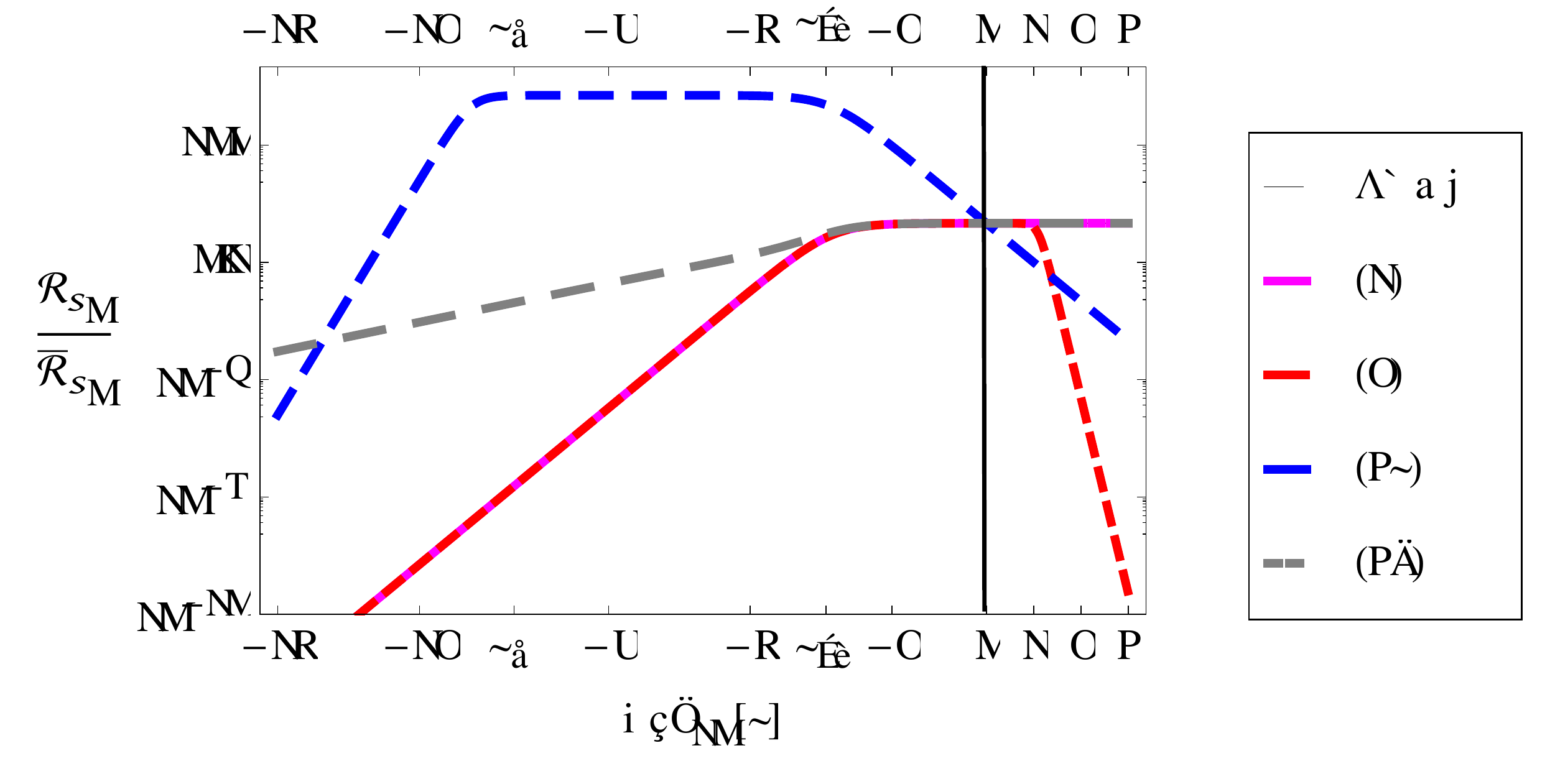}
\caption{Evolution of ${\cal R}_{{s_0}}/\bar{\cal R}_{{s_0}}$ for the four
  cases described in table \ref{tab:Leg1}. 
The curves for $\Lambda$CDM and case (1)  are overlapped.}
\label{Evol1}
\end{center}
\end{figure}
 \begin{figure}[h!!]
\begin{center}
\includegraphics[width=13cm,height=7cm]{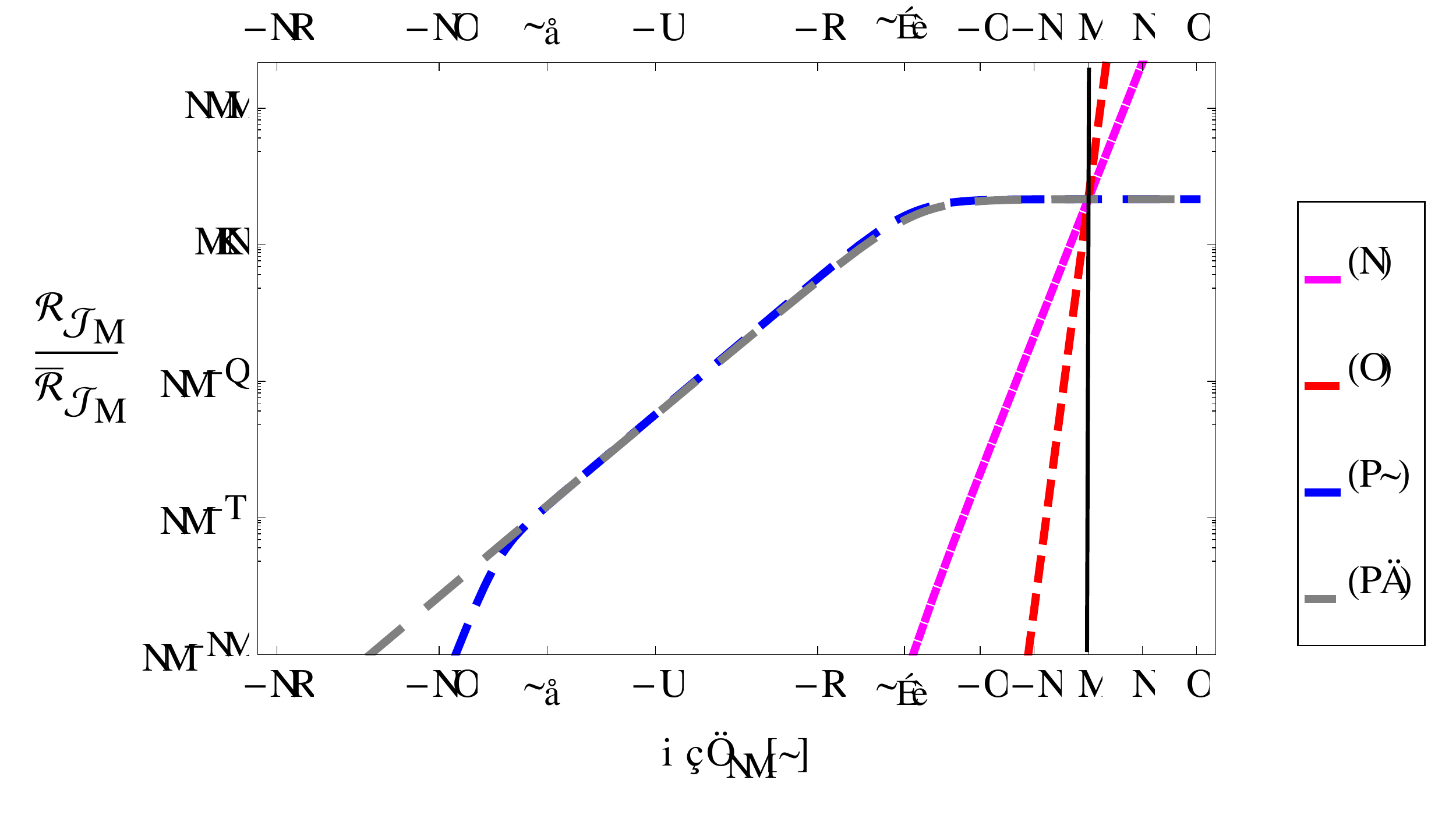}
\caption{ 
Evolution of the ratio ${\cal R}_{{\Im_0}}/\bar{\cal R}_{{\Im_0}}$ for the four
  cases described in table \ref{tab:Leg1}.}
\label{Evol2}
\end{center}
\end{figure}

\no
For case 1, we have  taken $\Omega_\Lambda=1-\Omega_r-\Omega_m$ as in the $\Lambda$CDM case.
   For case 2,  $ \Omega_\Lambda=0.7498$ and $\Omega_z=1-\Omega_r-\Omega_m-\Omega_\Lambda=10^{-4}$.
   For 3$a$ we have set $ \Omega_z=10^{-26}$ and $\Omega_\Lambda=1-\Omega_r-\Omega_m-\Omega_z$.
   Finally, for 3$b$  we have taken $ \Omega_z=3\;10^{-4}$ and $\Omega_\Lambda=1-\Omega_r-\Omega_m-\Omega_z$.
   \\
   Models 1 and 2 are characterised by an entropic active CC  (potentials containing terms proportional to the combinations $b\;Y$) 
 while for models 3 and 4 the  DeSitter phase is induce by a sterile CC.
\begin{itemize}
\item
 The first   model is characterised by a late time grow of the entropic perturbations, precisely from the period of DM-DE equality $a\sim (\Omega_\Lambda/\Omega_m)^{1/3}$. Intrinsic  pressure perturbations continue to growth also in the future as ${\cal R}\sim a^3 \;\Im_0$ while the relative one are exactly the same of the $\Lambda$CDM model and flatten to ${\cal R}\sim s_0/3$.
\item
 Also the second model is characterised  by a late time fastest grow of the entropic perturbations. The intrinsic fluctuations grow   as ${\cal R}\sim a^9 \;\Im_0$ crossing the non perturbative regime   in nearest future. The relative perturbations are exactly the same of the $\Lambda$CDM model but suddenly, in the future, will drop to zero as ${\cal R}\sim 1/a^5$.
\item
 The   model   3$a$ is characterised by the presence of a kineton phase ($w_z=1$) of the Universe in the very early time, before nucleosynthesis.  In order to satisfy all the necessary constraints, the parameter $\Omega_z$ has to be extremely small $\sim10^{-26}$.
  The intrinsic contribution grows until
 $a\sim (\Omega_z/\Omega_r)^{1/2}$ and it ``flattens" to ${\cal R}\sim a_{eq}\;\Im_0\;\Omega_m /\Omega_z$.
  Interestingly the relative contribution grows  very fast as  ${\cal R}\sim a^2\;\Omega_m/\Omega_z$ until $a\sim (\Omega_z/\Omega_m)^{1/3}$, it flats at  ${\cal R}\sim \frac{1}{a_{eq} }$ until equality time and then it decreases as  ${\cal R}\sim a_{eq}/a$.
 \item
  Finally,   model 3$b$  is characterised by the presence of a phase  with $w_z=1/4$, intermediate to matter and radiation.
   The relative entropic contribution grows  slowly up to equality time as   ${\cal R}\sim a^{1/4}\;\Omega_z/\Omega_r$ and then it stays constant as in $\Lambda$CDM.
   The intrinsic corrections grow until equality time up to ${\cal R}\sim (\Omega_m/\Omega_r)^{3/4}$
   after they decrease  as ${\cal R}\sim 1/a^{3/4}$.  
 \end{itemize}
 
 The main impact of a non trivial entropic component in a dark fluid is the presence of a   time dependent contribution to superhorizon scalar perturbations in the late time dynamics.
 In the above examples we tried to show a variety of possibilities.
 The above behaviours will impact mainly on the integrated Sachs-Wolfe  effect (ISW) 
 whose  temperature fluctuations  result from the differential
 redshift effect of photons climbing in and out of  a time evolving potential perturbations from last scattering surface to the present day:
  \ba
\left( \frac{\Delta T_\gamma}{T_\gamma}\right)_{(ISW)}=
2\;\int^{\tau_0}_{\tau_{dec}}\,d\eta\;\partial_\eta\Phi(\eta, \;\vec{x}=\eta\;\vec{n})
 \ea
 with $\vec{n}^2=1$.  
 When $\Phi$ is constant  (from \eq{defRm} imposing $\dot \Phi=0$ we get ${\cal R}=\frac{5+3\;w}{3\;(1+w)}\;\Phi$) the ISW effect is zero, as approximately it happens for adiabatic perturbations in $\Lambda$CDM model. It is clear from Figures (1) and (2) that the time behaviours  of possible entropic seeds for cosmological perturbations have to be carefully estimated and included to check the stability of the $\Lambda$CDM predictions against possible modifications of paradigms.

\section{Conclusions} 
\label{sect:conc}
 In the present paper we have used the effective field theory description
of perfect non-barotropic fluids  showing that  intrinsic entropic effects can be
relevant. One of the major advantages of the field description is the fact that gives
consistent non perturbative handle of the non-adiabatic part in the
pressure variation. In addition one can also build a thermodynamical
description of the fluid by relating field operators to basic
thermodynamical variables. 
Then one can relate the non
perturbative dynamics of vorticity and kinematical vorticity to the effective field theory approach.
Such dynamics can be extracted from a Lagrangian whose
structure is representing the free energy of the fluid. Inserting a
metric in the Lagrangian formalism allows to 
directly obtain the red intertwined dynamical of gravity
 and of the fluid, together with  its thermodynamical
  properties in the presence of gravity.
 \\
On the perturbative side one can  develop  the cosmological perturbations
around FLRW spacetime. The effective description is very powerful and
one can simply derive the evolution of the gravitational Bardeen potential, curvature perturbation ${\cal R}$ and  
uniform density surface curvature perturbation $\zeta$ setting the conditions for the violation of   the Weinberg theorem.
The entropy per particle perturbation $\delta \sigma$ is conserved and it acts just as a source
term. 
On the more phenomenological side we  have analysed  the impact of  a  dark
sector, taken as a single non-barotropic  perfect fluid that mimics both
DM and DE.
In order to keep things simply
enough to avoid numerical computations,  we consider a set of two-fluid models
 composed by radiation and an interacting dark sector, neglecting baryons.
Our results are compared against a simple version of $\Lambda$CDM model composed by two barotropic fluids (radiation plus dark matter) and a cosmological constant. 
From the field theory point of view, the presence of non trivial
entropy  emerges by the presence in the fluid Lagrangian $U$ of the
operator $Y$ that  can be identified with the dark sector temperature.
The cosmological-like equation of state $w=-1$ can also be produced by
a Lagrangian of the form $U(b\, Y)$.
This   means that  a De Sitter phase can be obtained not only with a
$ \Lambda$ term but also with a peculiar non-barotropic  fluid.
 Entropic perturbations, relative and intrinsic, are computed for a
 class of toy models that have very     different temporal
behaviour of perturbations triggered by entropic effects.
 Bounds on the scale of the corrections, proportional to $s_0$, for
 relative pressure perturbations, and to $\Im_0$ for intrinsic
 perturbations has to be settle by CMB \cite{Langlois:2012tm} 
and large scale  by the  power spectrum
\cite{Langlois:2000ar}. 
In this work we showed how powerful can be the EFT formalism for the
description of perfect fluid to study systematic way entropic effects  
and to simply recover  known results
on the non-perturbative dynamics of a fluid in the presence of
gravity.
 For the models studied,  we focused
 on the superhorizon evolution of the cosmological perturbations in presence of entropic fluid that can describe the yet unknown dark sector. 
 Late time growing corrections to ${\cal R}$ are potentially at work
 (ISW effect) for quite general dark fluids that mimic dark matter and
 later dark energy. A numerical investigation to  set the impact on
 Planck Data is necessary and  it will given in a separate paper.
 It would be interesting to play a similar  game  during inflation/reheating.

 \begin{appendices} 
 
 \section{  Single entropic fluid model as two barotropic fluids
   }
\label{section:ONE-baro}
It is interesting to investigate the behaviour of  a simple single fluid model which  reproduces the main features   of the $\Lambda$CDM model. 
Take the following potential (without the radiation fluid of photons/neutrinos)
\ba U=6 \, H_0^2 \, \plm^2 \;
\left(\frac{\Omega_r}{3}\;
  Y^{4}-\Omega_m\;b-\Omega_{\Lambda} \right) \, .
 \label{UL} 
\ea
The sound speed $c_s^2$ and the equation of state $w$ are the same of
$\Lambda$CDM and are given by (\ref{lcdm}). This it means that the background evolution from \eq{UL} and the $\Lambda$CDM model are the same.  Being $c_b^2=1/3$ constant, from (\ref{Tscal})  the temperature of such a fluid scale as  
$ 
T=Y=\frac{T_0}{a} \, , \; \phi' = \varphi_0=\text{const.} 
$ 
The evolution of $T$  is typical of  relativistic particles.
The evolution of ${\cal R}$ is driven, at superhorizon scales, by 
intrinsic non-adiabatic perturbations  (the relative entropic pressure is zero being the system composed by a single fluid)
with
 \be
\Gamma_{\text{tot}} = \Gamma_{\text{int}}=\frac{\delta \sigma _0 \, \varphi_0}{a^3 \left(4 \,a_{\text{eq}}+3
  \, a\right)} \, .
\ee
according  to (\ref{zetaeqm}) we get
\be
{\cal R}= \frac{  \, \delta \sigma _0 \, \varphi _0}{8 \, \plm^2 \, H_0^2
  \, \Omega_r  }\;\frac{a  }{  \left(3  \, a +4 \,  a_{\text{eq}} \right)} \, .
\label{Rr}
\ee
Interestingly, the  present single fluid model gives the very same
evolution for superhorizon curvature perturbations as in $\Lambda$CDM,
see \eq{SS}, upon the following identification 
\be
({\rm in} \; \Lambda{\rm CDM})\;\;s_0 \leftrightarrow - \frac{\delta \sigma_0 \, \varphi_0}{8 \, 
    H_0^2\, \plm^2 \, \Omega_r}\equiv \Im_0 \, .
\label{sident}
\ee
The dynamics of superhorizon perturbations of the single fluid model
with Lagrangian \eq{UL} is the same of the $\Lambda$CDM model, however
differences are present at small scales where for the $\Lambda$CDM model we have $v_d\neq 0$ while for 
the single fluid we have of course $v_d=0$.

The above result can be generalised to a single fluid described by 
\be
U=6\, \plm^2 \,  H_0^2\;  \left[\, w_r\;\Omega_r\; Y^{(1+w_r)/w_r} -\Omega_m\; b^{(1+w_m)}- \Omega_{\Lambda} \,\right],
\ee 
whose superhorizon perturbations are the same to the sum of two
perfect barotropic fluids with equation state of state
 $w_r$ and $w_m$ respectively.

\section{Details of Case 3}
\label{appcase1}
The full expression for ${\cal R}$ in case 3$a$ of section \ref{case1} is
\be
{\cal R} = \frac{a \left[\Omega_m \, a^{3 w_z} \left(
\delta \sigma
      _0+6 \, 
   H_0^2 \, \plm^2 \,  s_0 \, \left(w_z+1\right) \Omega _z \right)+6
 \, H_0^2 \, \plm^2 \,  s_0
   \left(w_z+1\right){}^2 \, \Omega _z^2\right]}{6 \, H_0^2 \plm^2 
   \left(w_z+1\right) \, \Omega _z \left[a^{3 \, w_z} \left(3 \, a \, \Omega
   _m+4 \, \Omega _r \right)+3 a \left(w_z+1\right) \Omega_z\right]} \, ;
\ee
\;
when $1/6<w_z <1/3$,  while for  $w_z >1/3$ we have (case 3$b$)
\be
{\cal R} =
\frac{a^{3 \, w_z} \left[a \, \delta \sigma _0 \, \Omega _m-8 \, H_0^2 \, \plm^2
    \,  s_0 \, 
   \Omega _r \left(w_z+1\right) \Omega _z \right]}{6 \, H_0^2 \,
 \plm^2 \left(w_z+1\right) \Omega _z \left[a^{3 \, w_z} \left(3 \, a
     \, \Omega_m+4 \, \Omega _r\right)+3 a \left(w_z+1\right) \Omega _z\right]} \, .
\ee
\end{appendices}

\bibliographystyle{JHEP}  
  
\bibliography{fluidbiblio}
  
\end{document}